\RequirePackage{fix-cm}
\documentclass[smallextended]{svjour3}       % 
\smartqed
\usepackage{amsmath}
\usepackage{graphicx}
\usepackage{cite}
\usepackage{xcolor}
\usepackage{amsfonts}
\usepackage[multiple]{footmisc}
\usepackage{graphicx}

  \usepackage[toc,page]{appendix}
  \makeatletter
\DeclareFontFamily{OMX}{MnSymbolE}{}
\DeclareSymbolFont{MnLargeSymbols}{OMX}{MnSymbolE}{m}{n}
\SetSymbolFont{MnLargeSymbols}{bold}{OMX}{MnSymbolE}{b}{n}
\DeclareFontShape{OMX}{MnSymbolE}{m}{n}{
    <-6>  MnSymbolE5
   <6-7>  MnSymbolE6
   <7-8>  MnSymbolE7
   <8-9>  MnSymbolE8
   <9-10> MnSymbolE9
  <10-12> MnSymbolE10
  <12->   MnSymbolE12
}{}
\DeclareFontShape{OMX}{MnSymbolE}{b}{n}{
    <-6>  MnSymbolE-Bold5
   <6-7>  MnSymbolE-Bold6
   <7-8>  MnSymbolE-Bold7
   <8-9>  MnSymbolE-Bold8
   <9-10> MnSymbolE-Bold9
  <10-12> MnSymbolE-Bold10
  <12->   MnSymbolE-Bold12
}{}

\let\llangle\@undefined
\let\rrangle\@undefined
\DeclareMathDelimiter{\llangle}{\mathopen}%
                     {MnLargeSymbols}{'164}{MnLargeSymbols}{'164}
\DeclareMathDelimiter{\rrangle}{\mathclose}%
                     {MnLargeSymbols}{'171}{MnLargeSymbols}{'171}
\makeatother
\usepackage[multiple]{footmisc}

\begin{document}

\title{Weak measurements, non-classicality and negative probability}
\author{Sooryansh Asthana$^*$         \and
        V. Ravishankar %etc.
}

  \institute{Sooryansh Asthana \at
              Department of Physics, Indian Institute of Technology Delhi, New Delhi-110016, India.\\ \email{sooryansh.asthana@physics.iitd.ac.in (Corresponding author)}           %  \\
           \and
           V. Ravishankar \at
              Department of Physics, Indian Institute of Technology Delhi, New Delhi-110016, India.\\
              \email{vravi@physics.iitd.ac.in}           
}

\date{Received: date / Accepted: date}

\maketitle

\begin{abstract}
  This paper establishes a direct, robust, and intimate connection among (i) non-classicality tests for various quantum features, e.g., non-Boolean logic, quantum coherence, nonlocality, quantum entanglement, quantum discord; (ii) negative probability, and, (iii) anomalous weak values.  It has been shown \cite{Adhikary18, Asthana20} that the nonexistence of a classical joint probability scheme gives rise to sufficiency conditions for  nonlocality, a nonclassical feature not restricted to quantum mechanics. The conditions for nonclassical features of quantum mechanics are obtained by employing  pseudo-probabilities, which are expectation values of the parent pseudo-projections. The crux of the paper is that the pseudo-probabilities, which can take negative values, can be directly measured as anomalous weak values.  We expect that this opens up new avenues for testing nonclassicality via weak measurements, and also gives deeper insight into negative pseudo-probabilities, which become measurable. A quantum game, based on violation of a classical probability rule, is also proposed that can be played by employing weak measurements. 
\end{abstract}

 %----------------------------------------------------------------
%\maketitle
  %----------------------------------------------------------------

\section{Introduction} 
\label{Introduction}
Since its very inception, quantum mechanics has been   looked upon in many ways, as may be seen from its different formulations  \cite{Schrodinger26, Heisenberg25, Dirac27, Feynman48, Bohm52, Nelson66}.  Its unique   features have also been  explored  and expanded in several  seminal works \cite{Einstein35,  Scrodinger35, Bell64, Kochen67}.  Presently, 
investigations of fundamental features of quantum mechanics have acquired an  importance like never before, for two reasons. Conceptually, 
there is an improved understanding of the so-called non-classical features of quantum states and, equally importantly,   there are new insights into the very concept of  the measurement-- due to weak measurements in quantum mechanics \cite{Aharonov88, Vaidman09}. In terms of practical impact, the importance  owes to non-locality, entanglement, quantum discord \cite{Bell64,  Horodecki09, Ollivier01} of states, which act as non-classical resources  for applications to quantum computation \cite{Knill98, Shi17},  and quantum information processing  \cite{Bennett92, Bennett93,  Ekert91, Streltsov12, Kanjilal18}, many of which are being implemented experimentally. 

This paper proposes to establish a robust connection between non-classicality,  as expressed through notions, such as non-Boolean logic \cite{Birkhoff36}, non-locality, entanglement, and quantum discord,  and the anomalous weak values,  as espoused in the concept of weak measurements. In doing so, the paper  aims to elucidate the key notions of quantum probability via weak measurements, through which the so-called negative probability becomes an observable quantity. 
Thereby, it is hoped that this study illustrates how the traditionally understood non-classical aspects of a quantum state and anomalous weak values are intimately intertwined with each other. 

We have two starting points.  The first one is the concept of nonclassical probability. It has been shown that various criteria for nonlocality and entanglement emerge as violations of classical probability rules\cite{Fine82, Adhikary18, Asthana20}. In quantum mechanics, pseudoprobability emerges as the  expectation value of a pseudo-projection, introduced recently
to explore non-classical features of quantum states directly in the language of quantum probability\cite{Adhikary18}. A special case,  which is also the simplest example,  is the Margenau-Hill-Barut distribution \cite{Hill61, Barut88} which serves our purpose in this paper.
 More generally, in principle -- by their very construction -- pseudo-projections capture  all possible criteria, such as the ones laid out in  \cite{Peres96, Horodecki96a, Ollivier01, Guhne03}. An important feature of this approach is that it does not employ involved algebraic techniques.  All the criteria of nonclassicality follow from violations of classical probability rules in a fairly straightforward manner. The task is to identify appropriate sums of pseudoprobabilities. Furthermore, pseudoprojections also capture violation of Boolean logic in the quantum domain.

 In fact, numerous entanglement inequalities\footnote{Entanglement inequalities are violated by all the  separable states and obeyed by at least one entangled state.}  emerge by imposing nonclassicality conditions on pseuodprobabilities \cite{Adhikary18, Asthana20}.  The criteria for nonlocality emerge when a few entries of a joint probability scheme turn negative, without recourse to any model. It is a manifestation of the fact that nonlocality is a nonclassical feature, not restricted to quantum mechanics.  In this way, negative probability not only provides us with a unified framework to obtain various criteria for different nonclassical features but also brings out the subtle difference between them.

The second starting point is the concept of weak measurements \cite{Aharonov88,  Vaidman08}, which has 
expanded the scope of measurements beyond the traditional projective measurements. Weak measurements, which admit experimental implementation,  display non-classicality by predicting anomalous weak values, which are otherwise   forbidden for projective measurements. In particular, the outcomes of
weak measurements may lie beyond the values allowed by the spectrum of the observable.  The relationships between anomalous weak values with  contextuality,  and  with counterfactual processes have been studied in \cite{Pusey14} and  \cite{Hosoya10} respectively. Of even greater relevance to us  are the weak measurements of noncommuting observables. Schemes for such measurements have been proposed \cite{Lundeen12}, and they  have already been implemented experimentally  \cite{Bamber14, Higgins15, Piacentini16}.  Recently, the weak value of local projection has been shown to appear as the modification in weak couplings  in interferometric alignment\cite{Dziewior19}.

 Equipped with the weak measurement techniques, we show how they can be employed to observe the violation of Boolean logic in the quantum domain. Furthermore, this paper undertakes the task of expressing  non-classicality tests, mainly based on the recent formulation by employing pseudo-projections \cite{Adhikary18},   in terms of weak measurements of appropriate projections, followed by a post-selection. In particular, we focus on non-locality, entanglement inequalities, and condition for discord in two-qubit systems.  This opens up new experimental avenues to test quantum features of states. Equally pertinently, we gain a better understanding of  the  negative pseudo-probability \cite{Adhikary18}  (as also of negative probabilities which were introduced much earlier, in a rather intuitive way, by  Dirac \cite{Dirac42} and Feynman \cite{Feynman87}, and  which was further made use of in \cite{Chandler92}).
Negative probabilities  are no more confined to the conceptual realm,  but  become measurable, as negative anomalous  weak values.

The paper is organised as follows. The next section (\ref{Preliminaries}) summarises  the formalisms employed for deriving conditions for non-Boolean logic, quantum coherence,  entanglement, quantum discord, and nonlocality. Section (\ref{Pseudo_weak}) presents the basics of weak measurements, to the extent needed for the work and  establishes the interrelation between pseudo-probabilities and anomalous weak values. In section (\ref{Notation}), we setup notations for an uncluttered discussion.  In section (\ref{Relation}), we discuss the relation of the present work with earlier works.  In section (\ref{Coherence}), anomalous weak value sufficient for a single qubit to be coherent has been identified. In section (\ref{Logic}),  we turn our attention to features of quantum logic that violate
classical Boolean logic directly, and, show that the weak measurements can be used to directly test those violations.  In section (\ref{Games}),  a quantum game,  based on violation of classical probability rules to be played employing weak measurements, has been given. The results  of the paper involving non-classical correlations are contained in section (\ref{Bell-CHSH}), (\ref{Ent_ineq}) and (\ref{Discord}). In these sections,  in the same order, we identify  the anomalous weak values that are sufficient for (i) the CHSH non-locality, (ii) families of entanglement inequalities for two-qubit systems, and, (iii) condition for discord (again in two-qubit systems). In section (\ref{Discussion_choice}), we have discussed how the hierarchical structure among different nonclassical correlations gets reflected in the choices of pseudoprobabilities. The last section (\ref{Conclusion}) summarises the results and discusses the scope for further work.

%----------------------------------------------------------------
\section{Preliminaries}
\label{Preliminaries}
In this section, we discuss our approach for deriving conditions for numerous nonclassical features.
\subsection{Non--Boolean logic, entanglement,  quantum discord, and quantum coherence}
Non-Boolean logic, quantum entanglement, quantum discord, and quantum coherence, etc. are nonclassical features of quantum mechanics, which is a new theory of probability. It does not follow Kolmogorov's axioms of classical probability\footnote{For example, in quantum mechanics, probability amplitudes add, and not the probabilities.}. We discuss here how, by employing standard rules of quantum mechanics, classical indicator functions for joint events map to pseudo-projection operators, and the  pseudoprobabilities in quantum mechanics can be defined as expectation values of the pseudoprojections.  The logical propositions involving conjunction (AND operation), negation (NOT operation), and disjunction (OR operation) also get represented by these pseudoprojection operators. 

In particular, we first recapitulate how Margenau-Hill distribution emerges from a  very specific  class of hermitian representatives of indicator functions for joint events.  For a detailed discussion, please refer to \cite{Adhikary17, Adhikary18}. 
\subsubsection{Joint probabilities and pseudo-projections}
\subsubsection*{ For a single observable}
Consider the event that  an observable $M$ takes a value $m$.
 For a classical system, the probability for the  event  can be determined by, (i) identifying the support for the  event in the phase space (and more generally, in the event space) and, (ii) finding the overlap of the indicator function for the support  with the given state (probability density). Recall that an indicator function is  a dichotomic  Boolean observable, taking  value 1 in the support and 0 elsewhere.
  In quantum mechanics,  the observable $M$ would be represented by a hermitian operator\footnote{For the sake of brevity, we represent both observables and operators by the same symbol throughout.} $M$
 in a Hilbert space, which
 admits the  eigen-resolution $M = \sum_i m_i \pi_{m_i}$.  The probability that the observable $M$ takes a value $m_i$, for a system in  a state $\rho$,  is given by the 
 overlap, ${\rm Tr}(\rho\pi_{m_i})$. The projections, $\pi_{m_i}$, are thus the  quantum representatives  of  the parent indicator functions. 
\subsubsection*{For two observables}
 Classically, the  indicator function for   joint outcomes of any two observables is simply the product of the respective indicator functions, which is non-vanishing only over the intersection of the two supports. However,  in quantum mechanics, observables do not necessarily commute, and there is no projection operator that would represent the classical indicator function for joint outcomes. That is,
 the indicator function for the intersection does not always map to a projection. Non-classicality can be understood by constructing their quantum representatives.

Thus, consider two observables $M, N$.  Let $\pi_{m_i},~\pi_{n_j}$
be the projection operators representing the respective indicator functions  for the outcomes $M = m_i$ and $N = n_j$.  The operator, representing the indicator function
 for the classical joint outcome, which we term as {\it  pseudo-projection} (PP), is given by the symmetrised product:
 \begin{equation}
 \label{PP}
 \mathbf{\Pi}_{m_i n_j} = \frac{1}{2}\left\{ \pi_{m_i},\pi_{n_j} \right\},
 \end{equation}
  in accordance  with the Weyl prescription \cite{Weyl27}.  
The PP is not idempotent,  unless $[\pi_{m_i},~\pi_{n_j}]=0$.  

In this way, the logical proposition, ${\cal L}(M=m_i)\wedge {\cal L}(N=n_j)$, gets represented by the pseudoprojection operator ${\bf \Pi}_{m_in_j}$ (the symbol $\wedge$ represents logical conjunction).

\subsubsection*{For multiple observables}
This correspondence suggests that the PP, representing joint outcomes of more  than two observables, can also be constructed similarly. It may be done so, but not in  a unique way. For, the order in which the noncommuting projections are multiplied matters.
Consider the  joint outcome,   $O_1 =o_1, O_2= o_2, \cdots, O_N=o_N$,  of $N$ observables. If $\pi_{o_i}$ be the respective 
projections,  their product can be permuted, in general, in $N!$ ways. If all the projections happen to be distinct, it will give rise to $N!/2$ distinct quantum representatives of the form,
 \begin{equation}
 \label{UnitPP}
{\bf \Pi}_{o_1\cdots o_N} = \frac{1}{2} \pi_{o_1}\pi_{o_2}\cdots\pi_{o_N}+ ~~{\rm h.c.},
 \end{equation}
where h.c. represents the hermitian conjugate.  A PP obtained from a given order  of $N$ distinct projections has been termed as a {\it unit pseudoprojection} \cite{Adhikary18} (That a unit PP has at least one  negative eigenvalue is shown in Appendix \ref{Eigen_conjunction}). Each unit PP  is, generally,  inequivalent to the others.  It is necessary to require that PPs share as many properties as possible with their classical counterparts. Unit PP fails one important test: symmetry in all the observables. This can be restored by summing all distinct unit PP with equal weights. But, not surprisingly, the resultant PP would be more difficult to analyse for nonclassicality. The most general PP is highly non-unique, and  may be taken to be   any  point on the manifold of  convex sums  of all unit PPs\footnote{For example, the most general PP representing the joint outcome, $O_1=o_1, O_2=o_2$ and $O_3=o_3$ is,
\begin{align}
{\bf \Pi}=\nu_1{\bf \Pi}_1+\nu_2{\bf \Pi}_2+\nu_3{\bf \Pi}_3;~0\leq \nu_i\leq 1;~\sum_{i=1}^3\nu_i=1,\nonumber
\end{align}
where, 
\begin{align}
{\bf \Pi}_{1}=\dfrac{1}{2}(\pi_{o_1}\pi_{o_2}\pi_{o_3}+{\rm h.c.});~{\bf \Pi}_{2}=\dfrac{1}{2}(\pi_{o_2}\pi_{o_1}\pi_{o_3}+{\rm h.c.});~{\bf \Pi}_{3}=\dfrac{1}{2}(\pi_{o_1}\pi_{o_3}\pi_{o_2}+{\rm h.c.}).\nonumber
\end{align}
}.  Fortunately,  for most of our purposes, the unit PP suffices.   

So, the  logical proposition, ${\cal L}(O_1=o_1)\wedge {\cal L}(O_2=o_2)\wedge \cdots\wedge {\cal L}(O_N=o_N)$, gets represented by different PPs depending on the prescription employed.
\subsubsection{Pseudoprobability}
\noindent We name the expectation of PP as pseudo-probability, i.e., the pseudoprobability for the joint event when the observables $M$ and $N$ take values $m_i$ and $n_j$ respectively, i.e., $M=m_i$ and $N=n_j$, is given by,
\begin{align}
{\cal P}_{m_in_j} \equiv {\rm Tr}({\bf \Pi}_{m_in_j}\rho).
\end{align}
When we employ a unit PP, the resultant pseudo-probability agrees with the Margenau-Hill-Barut distribution\cite{Hill61, Barut88}, the object of our study here.
 Since only the  hermiticity of PP  is guaranteed,  pseudo-probabilities can take negative values.   The PP for a multi-party system is the direct product of PP for individual systems. 

The set of pseudo-probabilities, generated for all the possible joint outcomes  of $N$ observables, is called a {\it scheme} \cite{Adhikary18}. 
Schemes possess an important property: the marginal, obtained by summing over the outcomes of any one of them, yields the scheme for the remaining $(N-1)$ observables. Importantly, the ultimate marginal, for a single observable,  is just the set of quantum probabilities, $p_k ={\rm Tr}(\rho \pi_k)$.   
\subsubsection{Definition of nonclassicality}
\label{Definition}
\noindent {\it Definition:} A state $\rho$ is deemed to be nonclassical with respect to a set of  observables $\{O_1, \cdots, O_N\}$, iff at least one pseudoprobability in the scheme assumes a negative value \cite{Adhikary18}.

If all the pseudo-probabilities were to be non-negative,  quantum mechanics would be admitting  an underlying classical description. That is to say, in an abstract manner, there does exist a classical system giving rise to the same set of nonnegative probabilities. This leads to the  attractive thesis  that pseudoprobabilities capture all the non classical features of quantum mechanics.  If it be so, one should be able to recover, in the first instance, features of nonclassical logic and, of course,  important results such as conditions for entanglement. That it is indeed so, has been shown recently \cite{Adhikary18, Asthana20}.

\subsubsection{Disjunction}
Conjunction is but one logical operation. The   PPs, that represent the 
disjunction (OR), may be obtained by employing standard rules. Thus, for example, the disjunction of two events when the observables $M$ and $N$ respectively take values $m_i$ and $n_j$, i.e.,  $M= m_i$ OR $N =  n_j$,  is represented by the PP,
\begin{equation}
\label{ORproj}
{\bf\Pi}_{m_i\vee n_j} = \pi_{m_i} + \pi_{n_j} - {\bf \Pi}_{m_in_j},
\end{equation}
which is simply the quantum representative of the indicator function for the union: $1_{S_i \cup S_j} = 1_{S_i} + 1_{S_j} - 1_{S_i \cap S_j}$.  Here $S_i, S_j$ and $S_{i}\cap S_j$ represent the supports for the events, $M=m_i$, $N=n_j$ and $(M=m_i$  AND $N=n_j)$ respectively.

\subsection{Nonlocality}
\label{Nonlocalityformalism}
Since nonlocality is a feature of nonclasicality that is not restricted to quantum mechanics \cite{Popescu94}, we assume a nonnegative joint probability scheme and show that violation of CHSH inequality is tantamount to the existence  of negative entries in the joint probability scheme.
%---------------------------------------------------------------------
\section{Pseudo-probabilities and weak values}
\label{Pseudo_weak}
The idea of negative probabilities has been around quite a while \cite{Dirac42, Feynman87, Chandler92}. But what exactly is the physical significance of pseudo-probabilities, so obtained by us, especially when they are negative?  We show that,  apart from being useful theoretical constructs,  they have an operational value, i.e., they can be realised in experiments as anomalous weak values. Conversely, the anomalous outcomes of weak measurements can be given an operator description in the language of PPs.  With this, it becomes possible to devise tests, which are distinct from standard correlation measurements,  for different manifests of non-classical correlations using weak measurements. This is the main content of the paper. With weak measurements, the experimental scope for testing quantum mechanics gets widened. Further, using weak measurements, the violation of Boolean logic can also be observed in the quantum domain.

The identification of  weak measurements that are appropriate to study pseudo-probability is quite straightforward. First, recall that 
the weak value of  an observable $A$ on a pre-selected state $|\psi\rangle$ and a post-selected state $|\phi\rangle$ is given by \cite{Aharonov88}, 
\begin{align}
\langle A_w \rangle^{|\phi\rangle}_{|\psi\rangle}  := \dfrac{\langle \phi|A|\psi\rangle}{\langle \phi|\psi\rangle}.
\end{align}
If $\langle A_w \rangle^{|\phi\rangle}_{|\psi\rangle}$  lies outside the range of eigenvalues of the observable $A$, the weak value is termed  {\it anomalous}. Thus, all imaginary values are automatically anomalous. Generalisation to mixed pre-selected state has been done in \cite{Wiseman02}. The weak value of an observable $A$ in the pre-selected state $\rho_1$ and post-selected state $\rho_2$ is given as,
\begin{align}
\langle A_w\rangle_{\rho_1}^{\rho_2}:= \dfrac{{\rm Tr}(\rho_2A\rho_1)}{{\rm tr}(\rho_2\rho_1)}.
\end{align}
An alternative interpretation of weak value of an observable as a robust property of a single pre-selected  and post-selected quantum system is discussed in \cite{Vaidman17}.

 It is pertinent to this work that many schemes have been proposed \cite{Lundeen12} for joint measurement of non-commuting observables. These techniques   facilitate experimental  determination of joint probability schemes (for non-commuting observables) \cite{Bamber14,  Higgins15, Piacentini16}. 

 Of relevance for us is the relation of pseudo-probabilities with anomalous weak values. Let  ${\bf \Pi}_{o_1\cdots o_N}$ be a unit PP for a joint event $O_1=o_1, O_2=o_2, \cdots, O_N=o_N$. Its associated pseudo-probability may be written, in terms of weak values as,

\begin{equation}
\label{mixed_weak}
\langle {\bf \Pi}_{o_1\cdots o_N} \rangle_{\rho} = {\rm Tr}(\rho \pi_{o_1}) \llangle {\pi}_{o_2} \cdots \pi_{o_N} \rrangle_{\rho}^{\rho_{o_1}},
\end{equation}
where, the notation  $\llangle~\rrangle$ -- which shall be used everywhere -- emphasises that $\llangle {\pi}_{o_2} \cdots \pi_{o_N} \rrangle_{\rho}^{\rho_{o_1}}$ represents the real part of the weak value of the product of the remaining $(N -1)$ projections. The   pre-selected state is $\rho$ and  the post-selected state, 
$\rho_{o_1}$. The state $\rho_{o_1}$ is obtained by suitable normalisation of $\pi_{o_1}$.

A further generalization to convex sums (including the symmetrised sum) is straightforward.

The weak value in the RHS of equation  (\ref{mixed_weak}) is that of a nonhermitian operator and merits more description. We  employ the  resolution  ${\pi}_{o_2} \cdots \pi_{o_N} \equiv  H - iJ$ in terms of its   hermitian and antihermitian parts, and state the condition for nonclassicality through the chain of mutual implications,
\begin{eqnarray}
\label{mip}
\langle {\bf \Pi}_{o_1\cdots o_N} \rangle_{\rho} < 0 & \iff &  \llangle {\pi}_{o_2} \cdots \pi_{o_N} \rrangle_{\rho}^{\rho_{o_1}} < 0  \nonumber \\
& \iff & \llangle H   \rrangle_{\rho}^{\rho_{o_1}} + {\rm Im} \{\langle J \rangle_{\rho}^{\rho_{o_1}}\} <0,
\end{eqnarray}
which displays the relationship between negative pseudo-probabilities and a combination   of the  weak values of the two hermitian observables. The symbol $\rho_{o_1}$ represents the state obtained from $\pi_{o_1}$ after suitable normalisation, i.e., if $\pi_{o_1}$ is a $d$-- dimensional projection operator,
\begin{align}
\label{Normalisation}
\rho_{o_1} \equiv \dfrac{1}{d}\pi_{o_1}.
\end{align}
For the special case of pseudoprojections representing  joint outcomes of only two observables,
\begin{eqnarray}
\label{mip1}
\langle {\bf \Pi}_{o_1o_2} \rangle  < 0 \iff   \llangle {\pi}_{o_2} \rrangle_{\rho}^{\rho_{o_1}}<0,
\end{eqnarray} 
which is expressed entirely in terms  of the anomalous weak values of a single hermitian operator.  Indeed, equation (\ref{mip1})  establishes a strong equivalence between negative pseudo-probabilities and anomalous weak values. This equivalence   forms   the basis for most of the applications discussed here.

 Note that the expectation value $\langle {\bf \Pi}_{o_1o_2} \rangle$ cannot exceed $+1$ and hence, only the negative anomalous weak value corresponds to negative pseudoprobability. It is due to the Cauchy-Schwarz theorem. The expectation value of ${\bf \Pi}_{o_1o_2}$ is upper bounded by $+1$ as shown below,
\begin{align}
|\langle{\bf \Pi}_{o_1o_2}\rangle_{\rho}|&=\dfrac{1}{2}\Big|\langle(\pi_{o_1}\pi_{o_2}+\pi_{o_2}\pi_{o_1})\rangle_{\rho}\Big|\nonumber\\
&\leq \dfrac{1}{2}(|\langle\pi_{o_1}\rangle_{\rho}|\cdot|\langle\pi_{o_2}\rangle_{\rho}|+|\langle\pi_{o_2}\rangle_{\rho}|\cdot|\langle\pi_{o_1}\rangle_{\rho}|)\leq 1.
\end{align} 
The same argument can be extended to the expectation value of a unit pseudoprojection representing joint outcomes of any number of observables.
\section{Notation}
\label{Notation}
 First, we introduce
 some notations to make the subsequent expressions  less cluttered:
\begin{enumerate}
\item We consider only dichotomic observables, with eigenvalues $\pm 1$. For simplicity, if $A$ be such an observable, the respective eigenprojections will be denoted by $\pi_A$ and $\pi_{\bar{A}}$, corresponding to eigenvalues $+1$ and $-1$. 
We also denote the associated states, obtained by suitable normalisation, by $\rho_A$ and $\rho_{\bar{A}}$, as shown in equation (\ref{Normalisation}).
\item \begin{enumerate}
\item The events, that the observable $A$ takes value $+1$ and $-1$, are compactly represented by ${\cal E}(A)$ and ${\cal E}(\bar{A})$ respectively, i.e.,
\begin{align}
{\cal E}(A) \equiv {\cal E}(A=+1);~ {\cal E}(\bar{A}) \equiv {\cal E}(A=-1).
\end{align}
\item The logical propositions, that the observable $A$ takes value $+1$ and $-1$, are compactly represented by ${\cal L}(A)$ and ${\cal L}(\bar{A})$ respectively.
\end{enumerate} 
\item The symbol $P(A) (P(\bar{A}))$ represents the quantum probability for outcome +1 ($-1$) of the observable $A$, i.e.,
\begin{align}
P(A) \equiv P(A=+1);~P({\bar{A}}) \equiv P(A=-1).
\end{align}
\item  The symbol $ {\cal P }(A_1A_2\cdots A_p\bar{A}_{p+1}\bar{A}_{p+2}\cdots \bar{A}_{p+q} )$ represents the pseudoprobability for
the outcomes of the first $p$ observables to be +1, and of the next $q$ observables to be $-1$, i.e.,
\begin{align}
&{\cal P }(A_1A_2\cdots A_p\bar{A}_{p+1}\bar{A}_{p+2}\cdots \bar{A}_{p+q} )\nonumber\\
&\equiv {\cal P}(A_1=+1,\cdots ,A_p=+1, A_{p+1}=-1,\cdots ,A_{p+q}=-1).
\end{align}
 \item Observables for subsystems  of a bipartite state are denoted by $A_i, B_i$, i.e.,  the observables $A_i$ and $B_i$ correspond to the first and second subsystems respectively.
\item Finally, since all the observables are dichotomic with outcomes $\pm 1$, we employ the following shorthand notation,
\begin{align}
& \mathcal{P}(A_1 = A_2= B_1 = B_2) \equiv \mathcal{P}(A_1 A_2; B_1 B_2) +  \mathcal{P}(\bar{A}_1 \bar{A}_2; \bar{B}_1 \bar{B}_2)\nonumber\\
=&\mathcal{P}(A_1=+1, A_2=+1; B_1=+1, B_2=+1) \nonumber\\
+&  \mathcal{P}(A_1=-1, A_2=-1; B_1=-1, B_2=-1).
\end{align}
\item For a  qubit,  we employ lowercase letters to represent the observables, i.e., $a_i  \equiv \vec{\sigma}_1\cdot\hat{a}_i$ and $b_j  \equiv \vec{\sigma}_2\cdot\hat{b}_j$. 
\end{enumerate}
\section{Relation with previous works}
\label{Relation}
A nonclassical joint probability distribution was first sought for noncommuting observables by simply taking the product of projection operators \cite{Kirkwood33}, which was further studied by Barut 
\cite{Barut57}. Note that this can lead to complex values. Independently, negative probabilities were introduced by Dirac \cite{Dirac42} and Feynmann \cite{Feynman87}, in a rather ad-hoc manner. Subsequently, Margenau and Hill \cite{Hill61} and Barut {\it et al.} \cite{Barut88} have constructed a  nonclassical joint probability distribution by employing a completely symmetrised product of two and three projections. The resulting joint probabilities are real but can take negative values. In contrast to those studies, the encompassing idea of pseudoprojection, as a quantum representative of product of indicator functions, has been introduced in \cite{Adhikary18}. Their expectation values reduce to  Margenau-Hill-Barut distribution for two observables, and also for three observables, for a specific choice of pseudoprojection. In this formalism, conditions for entanglement have been derived, both for two-qubit and multi-qubit systems, by imposing suitable nonlcassicality conditions \cite{Adhikary18, Asthana20}. Furthermore,  it has also been shown in \cite{Adhikary18, Asthana20} that conditions for nonlocality emerge when suitably chosen sums of joint probabilities turn negative, without recourse to any underlying model.

Coming to weak values, it has been shown by Johansen  \cite{Johansen_04, Johansen__04} that the Margenau-Hill-Barut distribution, for joint outcomes of two noncommuting observables, can be observed in experiments through weak measurements \cite{Aharonov88}. In fact, Margenau-Hill-Barut distribution, for joint outcomes of position and momentum ($x$ and $p$),  has also been experimentally observed \cite{Bamber14}. In parallel, anomalous weak values have been shown to be signatures of contextuality \cite{Pusey14} and violation of Laggett-Garg inequality \cite{Pan20}. What we accomplish in this paper is to identify the anomalous weak values underlying many other forms of  nonlcassicality, {\it viz.}, quantum coherence, CHSH nonlocality for a bipartite system, quantum entanglement, and quantum discord in two-qubit systems. Going further, we also show how weak measurements can be used for experimental demonstration of non-Boolean logic, which has not been considered hitherto. Our study, thus, shows that the nonclassical features also show violations of classical probability rules, that can be experimentally demonstrated through weak measurements.

%-----------------------------------------------------------------
\section{Applications - I}
\label{applications-II}
Non-classicality is a characteristic of all the quantum states, including the simplest of them all -- the single qubit. This is amply reflected in the definitions of both weak measurement and PPs.  In fact, a single qubit is quite a rich lab to test quantum mechanics. With this in mind, we propose three applications: a witness for coherence, a direct verification of non-Boolean nature of quantum logic, and a quantum game based on violation of classical probability rule. Weak  measurements act as experimental probes in all cases.
\subsection{Coherence witness for a single qubit and weak values}
\label{Coherence}  
In a given  basis, a state has coherence if it has nonvanishing   off-diagonal elements  \cite{Streltsov17}. Let  the state $\rho$ be expressed in the eigenbasis of, say,  ${\sigma}_z$. The PP,
\begin{eqnarray}
\label{Coher}
{\bf \Pi}_{a_1a_2} = \frac{1}{4} \Big({1} + \hat{a}_1\cdot\hat{a}_2 + \vec{\sigma}\cdot(\hat{a}_1 + \hat{a}_2)\Big),
\end{eqnarray}
  always has  a positive overlap with $\rho$ whenever $\rho$ is diagonal, so long as $(\hat{a}_1+\hat{a}_2)$  lies  in the plane orthogonal to $\hat{z}$.  On the other hand, if $\rho$ possesses coherence, 
 one may always  judiciously  choose $\hat{a}_1$ and $\hat{a}_2$, out of the plane   such that it has a negative overlap with ${\bf \Pi}_{a_1a_2}$. It has been elaborated in appendix (\ref{appcoherence}).
 
Weak measurements can act as detectors of coherence, since,
\begin{eqnarray}
& {\cal P}(a_1a_2) &\equiv \big\langle{\bf \Pi}_{a_1a_2}\big\rangle_{\rho}  = {\rm Tr}(\pi_{a_1}\rho) {\big\llangle} \pi_{a_2}\big\rrangle^{\rho_{a_1}}_{\rho},
\end{eqnarray}
which implies that,
\begin{eqnarray}
&  {\cal P}(a_1a_2)<0 &\implies {\big\llangle} \pi_{a_2}\big\rrangle^{\rho_{a_1}}_{\rho}<0.
\end{eqnarray}
The preselected and the postselected states are, respectively, $\rho$ and $\rho_{a_1}$. Note that, an anomalous weak value is necessary and sufficient for the expectation value of the witness to be negative.
 
%-----------------------------------------------------------------

 \subsection{Non-Boolean quantum logic and  weak values}
      \label{Logic}
    Recently, weak measurement techniques have been employed to experimentally observe violation of classical rules such as pigeon-hole paradox \cite{Aharonov16, Reznik20}. In this section, we show how such techniques facilitate experimental demonstration of violation of classical logic in quantum mechanics. 

Absurd propositions have null support in event spaces. Thus, their indicator functions  are identically zero. We construct two simple   examples which  illustrate how the quantum representatives of such absurd propositions do not vanish, and give rise to nonvanishing weak values.  They provide  a state-dependent and a state-independent test of violations of Boolean logic respectively.
   
   \subsubsection{State-dependent test of violation of Boolean logic}
     Consider  the logical proposition, ${\cal L}(a_1)\wedge {\cal L}(a_2)\wedge {\cal L}(\overline{a}_1)$. Since this proposition  is the conjunction of three  events, ${\cal E}(a_1=+1),~ {\cal E}(a_2=+1)$ and ${\cal E}(a_1=-1)$ (recall $a_1 \rightarrow \vec{\sigma}\cdot\hat{a}_1$), it has a null support. It is because the probability for the conjunction of the three  events is identically zero  classically,  as it involves conjunction of an event ${\cal E}(a_1=+1)$ with its negation ${\cal E}(a_1=-1)$.    But,  from the rules of quantum mechanics, we find  that there is exactly one  associated unit PP which  is non-vanishing \cite{Adhikary17}: 
\begin{equation}
\label{absurd}
{\bf \Pi}_{a_1a_2\overline{a}_1} =  \frac{1}{2} \Big\{ \pi_{a_1}\pi_{a_2}\pi_{\bar{a}_1} + \pi_{\bar{a}_1}\pi_{a_2}\pi_{a_1}\Big\} = \dfrac{1}{8}\vec{\sigma}\cdot\{\hat{a}_2-\hat{a}_1(\hat{a}_1\cdot\hat{a}_2)\}.
\end{equation}
Its  expectation for a given state-- the associated pseudo-probability-- is nonzero unless the state is completely mixed!   On the other hand, classical Boolean logic would demand that the joint probability, corresponding to the absurd event, ${\cal E}(a_1=+1)$ AND  ${\cal E}(a_2=+1)$ AND ${\cal E}(a_1=-1)$, is identically equal to zero, i.e.,
\begin{align}
    {\cal P}_{a_1a_2{\bar a}_1} =0 ~~{\rm (classically)}.
\end{align}
 Thus, ${\cal P}_{a_1a_2{\bar a}_1} \neq 0$, is a definite signature of violation of Boolean logic\footnote{ This situation can occur elsewhere as well. For example, nonlocality can be detected by both--inequalities and Hardy-type paradoxes \cite{Hardy93}.}.

  \subsubsection*{Experimental verification with weak measurements}  
Note that the form of Eq (\ref{absurd}) suggests  weak measurements of all the three projections appearing in it. This can be done by employing three pointers, with $\rho$ as the pre-selected,    and the completely mixed state as the post-selected state.   
 The Hamiltonian that couples the system to the pointers   may be  chosen to be,
      \begin{eqnarray}
      H_{\rm int} = g({\pi_{a_1}}\otimes P_{1x} +{\pi_{a_2}}\otimes P_{2y} + {\pi_{\overline{a}_1}}\otimes P_{3z}),
      \end{eqnarray}
     where,  $P_{1x}, P_{2y}, P_{3z}$ represent the momenta of the three pointers canonical to their positions $x_1, y_2, z_3$ respectively,  and $g$ is the common coupling constant. If the pointers are initially in the state,
     \begin{eqnarray}
    \exp(-x_1^2/2\sigma^2)\otimes \exp(-y_2^2/2\sigma^2)\otimes \exp(-z_3^2/2\sigma^2),
     \end{eqnarray} 
     the  correlation among  the three pointer readings $\langle x_1y_2z_3\rangle \propto \langle {\bf \Pi}_{a_1a_2\overline{a}_1}\rangle_{\rho}$. 

  \subsubsection{State-independent test of violation of Boolean logic}
 It may be inferred from equation (\ref{absurd}) that the pseudoprobability of the classically absurd joint event ${\cal E}(a_1a_2\bar{a}_1)$ would be state-dependent. Furthermore, the completely mixed state eludes violation of Boolean logic as the pseudoprobability ${\cal P}(a_1a_2\bar{a}_1)$ vanishes. This leads us to the question whether there exists a state-independent test of non-Boolean logic that can be experimentally demonstrated with weak measurement. The answer to this is in the affirmative and can be obtained by constructing the quantum representative of classically absurd proposition, ${\cal L}(a_1)\wedge {\cal L}(a_2)\wedge {\cal L}(\overline{a}_1)\wedge {\cal L}(\bar{a}_2)$. 
 
 The pseudoprojection, representing the event ${\cal E}(a_1a_2\bar{a}_1\bar{a}_2)$, is given by \cite{Adhikary17}:
 \begin{equation}
\label{absurd2}
{\bf \Pi}_{a_1a_2\overline{a}_1\overline{a}_2} =  \frac{1}{8} \Big\{ (\hat{a}_1\cdot\hat{a}_2)^2-\frac{1}{3}\Big\},
\end{equation}
whose overlap with any qubit state $\frac{1}{2}(1+\vec{\sigma}\cdot\vec{p})$ is independent of $\vec{p}$.
\subsubsection*{Experimental verification with weak measurements}  
The form of Eq (\ref{absurd2}) suggests the  weak measurements to be performed. This can be done by employing four pointers, with $\rho$ as the pre-selected state,    and the completely mixed state as the post-selected state.   
 The Hamiltonian that couples the system to the pointers   may be  chosen to be,
      \begin{eqnarray}
      H_{\rm int} = g({\pi_{a_1}}\otimes P_{1x} +{\pi_{a_2}}\otimes P_{2y} + {\pi_{\overline{a}_1}}\otimes P_{3z}+ \pi_{\overline{a}_2}\otimes P_{4x}),
      \end{eqnarray}
     where,  as before, $P_{1x}, P_{2y}, P_{3z}, P_{4x}$ represent the momenta of the four pointers canonical to their positions $x_1, y_2, z_3, x_4$ respectively,  and $g$ is the common coupling constant. If the pointers are initially in the state,
     \begin{eqnarray}
    \exp(-{x_1}^2/2\sigma^2)\otimes \exp(-{y_2}^2/2\sigma^2)\otimes \exp(-{z_3}^2/2\sigma^2)\otimes \exp{(-x_4^2/2\sigma^2)},
     \end{eqnarray} 
      the  correlation among  the four pointer readings $\langle x_1y_2z_3x_4\rangle$ is proportional to the pseudoprobability  $\langle {\bf \Pi}_{a_1a_2\overline{a}_1\overline{a}_2}\rangle_{\rho}$.  
     
      In these exceptional cases of absurd propositions, the negativity of the pseudo-probability is not a necessary condition for non-classicality. Rather, any nonzero value assumed by them is a signature of non-Boolean nature of quantum mechanics. 
     \subsubsection{Violation of the distributivity law in quantum mechanics}
     Boolean logic follows the distributivity law, i.e., the following two propositions are equivalent,
     \begin{align}
     \label{Distributivity}
     {\cal L}\big(a_1\wedge (a_2 \vee a_3)\big) \equiv {\cal L}(a_1 \wedge a_2)\vee {\cal L}(a_1 \wedge a_3).
     \end{align}
     It has been observed in \cite{Birkhoff36} that the distributivity rule is not obeyed in quantum mechanics. In what follows, we show that weak measurements can be employed to observe violation of the distributivity law. 
     
     The pseudoprojection representing the event in the LHS of equation (\ref{Distributivity}), i.e.,  $  {\cal E}\big(a_1\wedge (a_2 \vee a_3)\big)$ can be constructed as follows,
     \begin{align}
         \label{DistributivityLHS}
     {\bf \Pi}_{a_1\wedge (a_2 \vee a_3)} &\equiv \dfrac{1}{2}\{\pi_{a_1}, {\bf \Pi}_{a_2\vee a_3}\}= \dfrac{1}{2}\{\pi_{a_1}, \pi_{a_2}+\pi_{a_3}-{\bf \Pi}_{a_2a_3}\}\nonumber\\
        &= {\bf \Pi}_{a_1a_2} +{\bf \Pi}_{a_1a_3}-\dfrac{1}{2}\{\pi_{a_1}, {\bf \Pi}_{a_2a_3}\}.
     \end{align}
          The pseudoprojection representing the event in the RHS of equation (\ref{Distributivity}), i.e.,  $ {\cal E}(a_1 \wedge a_2)\vee {\cal E}(a_1 \wedge a_3)$ can be constructed as follows,
\begin{align}
     \label{DistributivityRHS}
{\bf\Pi}_{(a_1 \wedge a_2)\vee(a_1 \wedge a_3)} &= {\bf\Pi}_{a_1a_2}+{\bf \Pi}_{a_1a_3}-{\bf \Pi}_{a_1a_2a_1a_3}.
\end{align}
Subtracting equation (\ref{DistributivityRHS}) from equation (\ref{DistributivityLHS}), the following equation results:
\begin{align}
 {\bf \Pi}_{a_1\wedge (a_2 \vee a_3)} -{\bf\Pi}_{(a_1 \wedge a_2)\vee(a_1 \wedge a_3)}= {\bf \Pi}_{a_1a_2a_1a_3}- \dfrac{1}{2}\{\pi_{a_1}, {\bf \Pi}_{a_2a_3}\},
\end{align}
which is clearly non-vanishing. For example,  if we consider a state $\rho_{{\bar a}_1}$, lying in the null space of $\pi_{a_1}$, the expectation value of the second term $ \dfrac{1}{2}\{\pi_{a_1}, {\bf \Pi}_{a_2a_3}\}$ vanishes, i.e.,
\begin{align}
{\rm Tr}\Big(\rho_{{\bar a}_1}\dfrac{1}{2}\{\pi_{a_1}, {\bf \Pi}_{a_2a_3}\}\Big) =0.
\end{align}
The first term, $ {\bf \Pi}_{a_1a_2a_1a_3} = \frac{1}{2}(\pi_{a_2}\pi_{a_1}\pi_{a_3}+ {\rm h.c.})$, however has a nonzero expectation value\footnote{The  unit pseudoprojection ${\bf \Pi}_{a_1a_2a_1a_3}$ is given as, 
\begin{align}
{\bf \Pi}_{a_1a_2a_1a_3} =\frac{1}{2}(\pi_{a_2}\pi_{a_1}\pi_{a_1}\pi_{a_3}+\pi_{a_3}\pi_{a_1}\pi_{a_1}\pi_{a_2})= \frac{1}{2}(\pi_{a_2}\pi_{a_1}\pi_{a_3}+\pi_{a_3}\pi_{a_1}\pi_{a_2}). \nonumber
\end{align}
} with the state $\rho_{{\bar a}_1}$.
\subsubsection*{Experimental verification with weak measurements}
The pseudoprobabilities, $\langle {\bf \Pi}_{a_1a_2a_1a_3}\rangle$ and $\frac{1}{2}\big\langle\{\pi_{a_1}, {\bf \Pi}_{a_2a_3}\}\big\rangle$, can be determined with weak measurements and the difference between the two is a signature of violation of distributivity law in the quantum domain.

 %---------------------------------------------------------------
\subsection{Quantum games via pseudo-projections and weak measurements}
\label{Games}
 In this section, we propose  a quantum game, based on violation of rules of  classical probability, which may be implemented through weak measurements.  

Consider a qubit in the state,
 \begin{equation}
 \rho =\dfrac{1}{2}\big({1} + \vec{\sigma}\cdot\vec{p}\big),
 \end{equation}
 undergoing a unitary transformation induced by the Hamiltonian, 
 $
 H =-\dfrac{1}{2}\hbar \omega_L \sigma_z.
 $
 
The idea is to look at joint pseudo-probabilities  for simultaneous outcomes of two incompatible observables, $\vec{\sigma}\cdot \hat{m}$
and $\vec{\sigma}\cdot \hat{n}$. We wish to  exploit their  ``anomalous" nature, i.e., they can be  negative or the pseudoprobabilities for the complementary event may exceed one.
We choose, 
 \begin{eqnarray}
\label{obs}
 \hat{m} &= \cos \theta \hat{i} +\sin\theta \hat{j};~~\hat{n} = -\sin\theta \hat{i} + \cos\theta \hat{j}.
 \end{eqnarray}
 Employing the shorthand notations, 
 \begin{eqnarray}
 {\cal P}_1 &= {\cal P}_{++};~~ {\cal P}_2 = {\cal P}_{--};~~ {\cal P}_3 = {\cal P}_{+-};~~{\cal P}_4 = {\cal P}_{-+},
 \end{eqnarray}
 the  expression for the transition matrix, ${\cal T}$  for the pseudo-probabilities follows from their time dependence,  and is given by ($\omega_L=1$),
 \begin{eqnarray}
 \label{Transition}
\left(\begin{array}{c} {\cal P}_1{(t)} \\ {\cal P}_2{(t)} \\{\cal P}_3{(t)} \\{\cal P}_4{(t)} \end{array}\right) &= \dfrac{1}{4}
\left(\begin{array}{cccc}  1+2\cos t & 1-2\cos t & 1- 2\sin t & 1+ 2\sin t\\ 1-2\cos t & 1+2\cos t & 1+2\sin t & 1-2\sin  t \\ 1+ 2\sin t & 1-2\sin t & 1+2\cos t & 1-2 \cos t \\1-2\sin t & 1+2\sin t & 1-2\cos t & 1+2\cos t\end{array}\right)
\left(\begin{array}{c} {\cal P}_1{(0)} \\ {\cal P}_2{(0)} \\{\cal P}_3{(0)} \\{\cal P}_4{(0)} \end{array}\right) \equiv {\cal T}(t)\left(\begin{array}{c} {\cal P}_1{(0)} \\ {\cal P}_2{(0)} \\{\cal P}_3{(0)} \\{\cal P}_4{(0)} \end{array}\right). \nonumber\\
\end{eqnarray}
${\cal T}(t)$ is an example of the quantum counterparts of the doubly stochastic matrices in classical probability theory. We note that ${\cal T}(t)$ satisfies all the properties of doubly stochastic matrix, except that its entries admit  negative values. The continuous set, \{${\cal T}(t)$\}, forms a monoid -- a semi group with identity. 

 If an entry in ${\cal P}{(t)}$ were to become negative, the sum of the other three entries would exceed one. This affords advantages which are not provided classically, and will be employed in the game. 
The rules of the game, that we design, are as follows:

\begin{itemize}
\item There are two players. The referee asks them  to start and end the game at definite times, $t_i$ and $t_f$, respectively. They are given the same  initial {\it non-negative}  scheme for joint outcomes of a set of two dichotomic observables. The players are to compete  to create, by unitary evolution,    the scheme such that at  some  intermediate time $T$,   the sum $S$, of any of its three entries maximally exceeds one. Weak measurements are to be employed.
\item The players are free to choose the initial  state and the Hamiltonian. 
\item In this game, a resource is classical, if for all times,     ${\cal T}(t)_{ij} \geq 0$.
\end{itemize}

We provide an explicit illustration of existence of such a pseudo-probability scheme that can be used as a strategy by players.
Suppose that  $\hat{m} = \hat{x},~\hat{n} = \hat{y}$,  and  the initial state (at $t=0$) is  pure with its polarisation along the $x$ axis.  The associated initial pseudo-probabilities are, 
\begin{eqnarray}
{\cal P}_1(0) = 0.5;~ {\cal P}_2(0) = 0;~ {\cal P}_3(0) = 0.5;~ {\cal P}_4(0) = 0.\nonumber
\end{eqnarray}
  At $T = \dfrac{\pi}{4}$, the pseudo- probabilities satisfy the maximality requirement,  and are given by,
\begin{eqnarray}
 & {\cal P}_1{(T)}=0.25;~ {\cal P}_2{(T)}=0.25;~\nonumber\\
 & {\cal P}_3{(T)}=0.25(1+\sqrt{2});~ {\cal P}_4{(T)}=0.25(1-\sqrt{2}),
\end{eqnarray}
which lead to the maximal value $S = 0.25(3 +\sqrt{2})$.
  If a player were to employ a different state or a different Hamiltonian,    the game would be lost.

\section{Applications - II}
\label{applications-I}
 The definition of nonclassicality, given in section (\ref{Definition}), is broad enough to detect all the quantum states nonclassical\footnote{Even the completely mixed two-dimensional state has a negative pseudo-probability for the joint event, when $\vec{\sigma}\cdot\hat{a}_1, \vec{\sigma}\cdot\hat{a}_2, \vec{\sigma}\cdot\hat{a}_3$ take value $+1$, where $\hat{a}_1, \hat{a}_2, \hat{a}_3$ are coplanar and at an included angle of $\frac{2\pi}{3}$. The completely symmetrised PP can be constructed as follows:
\begin{align}
{\bf \Pi}_{a_1a_2a_3}=\dfrac{1}{3!}(\pi_{a_1}\pi_{a_2}\pi_{a_3}+\pi_{a_1}\pi_{a_3}\pi_{a_2}+\pi_{a_2}\pi_{a_1}\pi_{a_3}+\pi_{a_2}\pi_{a_3}\pi_{a_1}+\pi_{a_3}\pi_{a_1}\pi_{a_2}+\pi_{a_3}\pi_{a_2}\pi_{a_1})=-\dfrac{1}{16},\nonumber
\end{align} 
whose overlap with the completely mixed state is negative, $-\frac{1}{16}$.}. From a  resource-theoretic viewpoint, however, specific non-classical features possessed by a state determine the applications, in which it will be useful. This aspect has led to various criteria for detection of those nonclassical features. In order to show that nonclassical probabilities and anomalous weak values underlie them, we start with various non-classical features, e.g., nonlocality, entanglement and discord. We pick up such sums of pseudo-probabilities that the ensuing inequalities get violated by  at least one state having that particular non-classical feature. We consider bipartite systems since their nonclassical properties  have received the greatest attention. 
\subsection{CHSH nonlocality and weak values}
\label{Bell-CHSH}
 In this section, we show how the derivation of CHSH inequality, in terms of joint probability given in \cite{Adhikary18}, facilitates identification of anomalous  weak values that guarantee a state to be non-local. 

Consider any  $d_1\times d_2$  dimensional bipartite system. Let $\{A_1, A_2\}, \{B_1, B_2\}$ be two pairs of dichotomic observables (with outcomes, $\pm 1$), for the respective  subsystems.    Since CHSH inequality involves terms containing only correlations between the observables of the two subsystems, and not the local terms, we choose the  joint probabilities in such a way that all the local terms get cancelled. 

The relevant joint probability is the sum,
\begin{equation}
\label{Nloc}
{\cal P}_{\rm NL}={\cal P}(A_1=B_1=B_2) +{\cal P}(A_2=B_1=\overline{B}_2).
\end{equation}
 We rewrite all the joint probabilities in terms of expectation values of observables, by employing dichotomic nature of observables.  Classically, it is always true that ${\cal P}_{\rm NL} \ge 0$. Therefore,  we  demand   that ${\cal P}_{\rm NL}<0 $, and  obtain the CHSH inequality,
\begin{eqnarray}
\label{BI}
\big\langle A_1 (B_1 + B_2) + A_2 (B_1-B_2)\big\rangle > -2.
\end{eqnarray}
The detailed derivation is shown in appendix (\ref{Belldetailed}).

In quantum mechanics, of course, the joint probabilities can be written as expectation values of the parent peudo-projections. Thus, in quantum mechanics, given the relation between pseudoprobabilities and anomalous weak values, the former can be determined via weak measurements. What would be the corresponding  weak measurements? We rewrite the RHS of the equation  (\ref{Nloc}) as,
\begin{eqnarray}
{\cal P}_{\rm NL} &  = & \langle \pi_{A_1}\pi_{B_1}\rangle 
\llangle \pi_{B_2} \rrangle_{\rho}^{\rho_{A_1B_1}} +
\langle \pi_{\bar{A}_1}\pi_{\bar{B}_1}\rangle 
\llangle \pi_{\bar{B}_2} \rrangle_{\rho}^{\rho_{\bar{A}_1\bar{B}_1}} \nonumber \\
\label{Bellpp}
& + & 
\langle \pi_{A_2}\pi_{B_1}\rangle 
\llangle \pi_{\bar{B}_2} \rrangle_{\rho}^{\rho_{A_2B_1}}+
\langle \pi_{\bar{A}_2}\pi_{\bar{B}_1}\rangle 
 \llangle \pi_{B_2} \rrangle_{\rho}^{{\rho}_{\bar{A}_2\bar{B}_1}},
\end{eqnarray}
where,
\begin{align}
\pi_{A} = \frac{1}{2}(1+A); \pi_{\bar A} = \frac{1}{2}(1-A);~~\rho_{AB} = \frac{1}{d_{\pi_{A}}d_{\pi_B}}\pi_A\pi_B,
\end{align}
 where $d_{\pi_{A}} (d_{\pi_B})$ represents dimensions of$\pi_{A} (\pi_B)$.

 Note that, if ${\cal P}_{\rm NL} < 0$,    at least some of the weak values are negative. In fact,  for two-qubit nonlocal Werner states, it is necessary that all the four weak values be negative (shown in appendix (\ref{Belldetailed})). With hindsight, we realise that 
for pure $2 \otimes 2$ systems, the pseudo-probabilities underlying CHSH inequality have already been experimentally demonstrated in \cite{Higgins15}. 

Our analysis also resolves the apparent paradox  posed in problem 17.6 of \cite{Aharonov08}. The paradox is that apparently, a normalised joint probability distribution exists reproducing all the desired marginals, even when the CHSH inequality gets violated. The resolution  is that,  though a joint probability distribution  does exist for  the four observables $\{A_i, B_j\}$,  a violation of  CHSH inequality requires that the weak values be  anomalous.  Thus, if equation (\ref{Bellpp}) assumes a negative value, it simultaneously signifies that,
\begin{enumerate}
\item the underlying state is CHSH nonlocal, 
\item the pseudo-probabilities become negative,  and, 
\item the weak values are anomalous.
\end{enumerate}
 In short, we have shown the existence of anomalous weak values and negative probabilities upon violation of CHSH nonlocality.   In sections (\ref{Ent_ineq}) and (\ref{Discord}), we show that similar conclusions hold for entanglement and discord in two-qubit systems as well.

\subsection{Entanglement inequality for two-qubits and weak values}
\label{Ent_ineq}
We next study the interrelation between entanglement inequalities, negative pseudoprobabilities and anomalous weak values. As mentioned in section (\ref{Introduction}), entanglement inequalities are  so formulated that the ensuing inequalities get satisfied by some entangled states and are violated by all the separable states.  In order to illustrate the interrelation,  we take up  two-qubit systems for our study. The same approach can be advanced to multi-qubit as well as to multi-qudit systems, by writing the entanglement inequalities as a sum of PPs with nonnegative weights (by employing the derivation given in \cite{Asthana20}). We first derive a set of two entanglement inequalities, which get expressed as linear combinations of pseudo-probabilities.
%Again, we employ unit pseudo projections which lead to a Margenau-Hill distribution. 
We first list, following   \cite{Adhikary18}, the criteria  for  choosing  pseudo-probabilities that lead to entanglement inequalities.
\subsubsection{Rationale underlying choice of pseudoprobabilites yielding entanglement inequalities}
\label{Rationale_ent}
\begin{enumerate}
\item Entanglement  in a two-qubit system refers to correlations between outcomes of one set of locally non-commuting observables on the first qubit with another set of locally non-commuting observables on the second qubit. Guided by this, we choose pseudo-probabilities for joint outcomes of  sets of observables,  which ensure that the ensuing inequality involves the correlations, such as  $\langle \vec{\sigma}_1\cdot\hat{a}\vec{\sigma}_2\cdot\hat{b}\rangle$ and $\langle \vec{\sigma}_1\cdot\hat{a}^{\prime}\vec{\sigma}_2\cdot\hat{b}^{\prime}\rangle$, with the proviso that the commutators $[\vec{\sigma}_1\cdot\hat{a}, \vec{\sigma}_1\cdot\hat{a}^{\prime}] \neq 0; [\vec{\sigma}_2\cdot\hat{b}, \vec{\sigma}_2\cdot\hat{b}^{\prime}] \neq 0$ (please note that the observables $\vec{\sigma}_1\cdot\hat{a}, \vec{\sigma}_1\cdot\hat{a}'$ and $\vec{\sigma}_2\cdot\hat{b}, \vec{\sigma}_2\cdot\hat{b}'$ act over the spaces of the first and the second qubit respectively.).

\item A further requirement is, of course, that, for separable states, the inequality that follows from non-classicality requirement gets automatically violated.
\end{enumerate}
\subsection*{Geometry of observables}
\begin{figure}[!htb]
        {\includegraphics[width=85mm,scale=1]
        {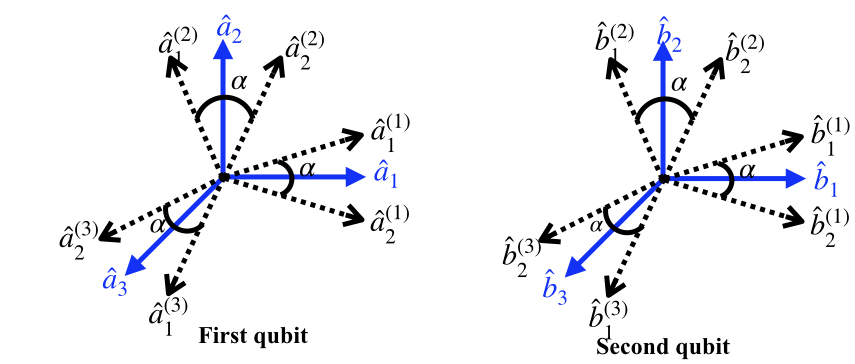}}
       \caption{$\hat{a}_1, \hat{a}_2, \hat{a}_3, \hat{b}_1, \hat{b}_2, \hat{b}_3$: Directions for first and second qubit  respectively that appear in the final entanglement inequality (shown in blue). \\$\{\hat{a}_i^{(1)}, \hat{a}_i^{(2)}, \hat{a}_i^{(3)}\}, \{\hat{b}_i^{(1)}, \hat{b}_i^{(2)},  \hat{b}_i^{(3)}\}, i \in \{1 ,2\}:$ Directions for the first and the second qubit involved in the construction of pseudo-projections (shown in black).}
         \label{nl_3fig}
      \end{figure}

The observables employed for constructing entanglement inequalities obey a common geometry, which is as follows:
(i) For each qubit, we have three sets  of doublets, 
   $\{ a^{(i)}_1, a^{(i)}_2\}$ and $\{ b^{(i)}_1, b^{(i)}_2\}$, $i =1,2,3$. Recall that generically, $a \equiv \vec{\sigma}_1\cdot \hat{a}$, $b \equiv \vec{\sigma}_2\cdot \hat{b}$. (ii) The angles  between the  two observables in each of  the six sets  have the same value,   which  is represented by $\alpha$, the only free parameter. (iii) For each qubit, the normalised sums of vectors from within each set forms an orthonormal basis, and are denoted by $\{\hat{a}_i\}$ and $\{\hat{b}_i\}$ respectively. Thus, $\vert\hat{a}_1\cdot(\hat{a}_2\times\hat{a}_3)\vert =\vert\hat{b}_1\cdot(\hat{b}_2\times\hat{b}_3)\vert = 1$. 
The geometry is completely depicted in figure (\ref{nl_3fig}).

\subsubsection{Linear entanglement inequalities}
 The  set of entanglement inequalities which we set forth  are, essentially, refinements over the construction given in \cite{Adhikary18}. We,  then, recast it  in the language of weak measurements. 
\subsubsection*{Inequality I}
\label{First_linear}
Of interest to us is the following sum of pseudo-probabilities,
\begin{equation}
\label{IneqI}
\mathcal{P}_{E_1} = \sum_{i=1}^2  \mathcal{P}(a_i = b^{(i)}_1 = b^{(i)}_2).
\end{equation}
Classical probability mandates the sum to be non-negative. Hence, for non-classicality, $\mathcal{P}_{E_1} < 0 $, which yields the criterion,
  \begin{equation}
  \label{Ent2qubit}
    2 \cos\frac{\alpha}{2}+\sum_{i=1}^2 
 \Big\langle  \vec{\sigma}_{1}\cdot\hat{a}_i\vec{\sigma}_{2}\cdot\hat{b}_i  \Big\rangle < 0.
\end{equation}
The necessary condition that,  $\mathcal{P}_{E_1} >  0$,  for all separable states is obeyed provided that   $0 < \alpha \le \frac{2\pi}{3}$.   The detailed derivation is given in appendix (\ref{Entlinapp_1}).

 In terms of weak values, Eq. (\ref{IneqI}) has the form,
\begin{equation}
\label{Ineq1weak}
\mathcal{P}_{E_1} =\sum_{i=1}^2\Big\{ \langle \pi_{a_i}\pi_{b^{(i)}_2}\rangle \llangle \pi_{b^{(i)}_1}\rrangle_{\rho}^{ \rho_{a_i}\rho_{b^{(i)}_2}}+ \langle \pi_{\overline{a}_i}\pi_{\overline{b}^{(i)}_2}\rangle \llangle \pi_{\overline{b}^{(i)}_1}\rrangle_{\rho}^{\rho_{\overline{a}_i}\rho_{\overline{b}^{(i)}_2}} \Big\},
\end{equation}
from which we see that four weak measurements are required. Also, the inequality (\ref{Ent2qubit}) can not get satisfied
 if {\it none} of the four weak values in (\ref{Ineq1weak}) is anomalous. For  the Werner states, detected to be entangled by inequality (\ref{Ent2qubit}), {\it all} the four weak values are anomalous, as shown in appendix (\ref{Entlinapp_1}).

\subsubsection*{Inequality II}
\label{Second_linear}
The relevant pseudo-probability is  the sum  \cite{Adhikary18},
\begin{equation}
\label{IneqII}
\mathcal{P}_{E_2}= \sum_{i=1}^3  \mathcal{P}(a_i = b^{(i)}_1 = b^{(i)}_2),
\end{equation}
which,  again, is non-negative classically. Hence, 
 the demand that, ${\mathcal{P}_{E_2}} < 0$,   yields the inequality,
\begin{equation}
\label{Ent2qubit1}
  3 \cos\frac{\alpha}{2} +\sum_{i=1}^3 \Big\langle \vec{\sigma}_{1}\cdot\hat{a}_i\vec{\sigma}_{2}\cdot\hat{b}_i \Big\rangle < 0. 
    \end{equation}
    The further requirement that, $ \mathcal{P}_{E_2} \ge 0$, for all separable states imposes the  additional constraint,  $0 < \alpha \le  \arccos(-7/9) \simeq \pi$.  The detailed derivation is given in appendix (\ref{Entlinapp_2}).

 In terms of weak values, $\mathcal{P}_{E_2} <0$ translates to the condition,
   \begin{eqnarray}
\label{Ineq2weak}
 {\cal P}_{E_2} & =  &  \sum_{i=1}^3\Big\{ \langle \pi_{a_i}\pi_{b^{(i)}_2}\rangle \llangle \pi_{b^{(i)}_1}\rrangle_{\rho}^{ \rho_{a_i}\rho_{b^{(i)}_2}} 
+  
 \langle \pi_{\bar{a}_i}\pi_{\bar{b}^{(i)}_2}\rangle \llangle \pi_{\overline{b}^{(i)}_1}\rrangle_{\rho}^{\rho_{\bar{a}_i}\rho_{\bar{b}^{(i)}_2}} \Big\} < 0,
\end{eqnarray}
which again demonstrates that 
 $ \mathcal{P}_{E_2} \ge 0$   if none of  the weak values is anomalous.
 In fact,   for  entangled Werner states, {\it all} the six weak values are required to be anomalous, as shown in appendix (\ref{Entlinapp_2}). 
 
 \subsubsection{Nonlinear Entanglement Inequalities}
 We have, so far, considered violation of classical probability rules for sums of chosen pseudo-probabilities. In this section, we explore violations of classical probability rules for bilinears in pseudo-probabilities and show how they can be harnessed to yield non-linear entanglement inequalities.
 %----------------------------------------------------------------
 
 \subsubsection*{Inequality I}
For construction of first nonlinear entanglement inequality, the combination of pseudo-probabilities  is given by, 
 \begin{equation}
 \label{nl1}
{\cal S}_1 =  \sum_{i=1}^2\mathcal{P}(a_i=b^{(i)}_1=b^{(i)}_2)\mathcal{P}(\overline{a}_i =b^{(i)}_1=b^{(i)}_2).
 \end{equation}
 Imposing the  non-classicality condition, ${\cal S}_1 < 0$,  we arrive at the inequality,
 \begin{equation}
 \label{nl_1}
 2\cos^2\dfrac{\alpha}{2} -\sum_{i=1}^2 \langle\vec{\sigma}_1\cdot\hat{a}_i\vec{\sigma}_2\cdot\hat{b}_i\rangle^2 <0.
 \end{equation}
 The further necessary condition that, ${\cal S}_1 \ge 0$,  for  all separable states yields the range $0 < \alpha \le \dfrac{\pi}{2}$.   The detailed derivation is given in appendix (\ref{Entnlinapp_1}).
 
  It remains to express Eq. (\ref{nl1}) in terms of weak values, which can be verified to be  given by,
\begin{eqnarray}
 \sum_{i=1}^2\Big\{ &\llangle \pi_{b^{(i)}_1}\rrangle_{\rho}^{\rho_{a_i}\rho_{b^{(i)}_2}} \llangle \pi_{b^{(i)}_1}\rrangle_{\rho}^{\rho_{\overline{a}_i}\rho_{b^{(i)}_2}} \langle \pi_{a_i}\pi_{b^{(i)}_2}\rangle\langle \pi_{\overline{a}_i}\pi_{b^{(i)}_2}\rangle+\nonumber\\
  \label{nl1weak}
& \llangle \pi_{\overline{b}^{(i)}_1}\rrangle_{\rho}^{\rho_{\overline{a}_i}\rho_{\overline{b}^{(i)}_2}}\llangle \pi_{\overline{b}^{(i)}_1}\rrangle_{\rho}^{ \rho_{a_i}\rho_{\overline{b}^{(i)}_2}}  \langle \pi_{\overline{a}_i}\pi_{\overline{b}^{(i)}_2}\rangle\langle \pi_{a_i}\pi_{\overline{b}^{(i)}_2}\rangle \Big\}.
\end{eqnarray}

Eq. (\ref{nl1}) cannot  acquire a negative value  if {\it none} of the eight weak values in (\ref{nl1weak}) is anomalous. If a set of four weak values, $\Big(\llangle \pi_{b^{(i)}_1}\rrangle_{\rho}^{\rho_{a_i}\rho_{b^{(i)}_2}}$  and $\llangle  \pi_{\overline{b}^{(i)}_1}\rrangle_{\rho}^{\rho_{\overline{a}_i}\rho_{\overline{b}^{(i)}_2}}\Big)$, or $\Big(\llangle \pi_{b^{(i)}_1}\rrangle_{\rho}^{\rho_{\overline{a}_i}\rho_{b^{(i)}_2}}$ and $\llangle \pi_{\overline{b}^{(i)}_1}\rrangle_{\rho}^{ \rho_{a_i}\rho_{\overline{b}^{(i)}_2}}\Big)$, are anomalous, the ensuing inequality (\ref{nl_1}) is bound to get satisfied.

 %----------------------------------------------------------------
   \subsubsection*{Inequality II}
  The bilinear combination of pseudo-probabilities is chosen to be,
 
 \begin{equation}
 \label{nl2}
 {\cal S}_2 = \sum_{i=1}^3\mathcal{P}(a_i = b^{(i)}_1 = b^{(i)}_2)\mathcal{P}(\overline{a}_i = b^{(i)}_1 = b^{(i)}_2), 
 \end{equation}
  which, following the same method, yields the inequality for the non-classicality condition ${\cal S}_2 <0$ to be,
  \begin{equation}
   \label{E_nl_2}
 3\cos^2\dfrac{\alpha}{2} -\sum_{i=1}^3 \langle\vec{\sigma}_1\cdot\hat{a}_i\vec{\sigma}_2\cdot\hat{b}_i\rangle^2 <0.
 \end{equation}
As usual, we demand that, ${\cal S}_2 \ge 0$, for all  separable states which imposes the constraint 
   $0 < \alpha \le  \arccos(-1/3)$. The detailed derivation is given in appendix (\ref{Entnlinapp_2}).

 Finally, the inequality (\ref{nl2}), in terms of weak values, has the form,
 \begin{eqnarray}
 \sum_{i=1}^3\Big\{ &\llangle \pi_{b^{(i)}_1}\rrangle_{\rho}^{\rho_{a_i}\rho_{b^{(i)}_2}} \llangle \pi_{b^{(i)}_1}\rrangle_{\rho}^{\rho_{\overline{a}_i}\rho_{b^{(i)}_2}} \langle \pi_{a_i}\pi_{b^{(i)}_2}\rangle\langle \pi_{\overline{a}_i}\pi_{b^{(i)}_2}\rangle+\nonumber\\
  \label{nl2weak}
& \llangle \pi_{\overline{b}^{(i)}_1}\rrangle_{\rho}^{\rho_{\overline{a}_i}\rho_{\overline{b}^{(i)}_2}}\llangle \pi_{\overline{b}^{(i)}_1}\rrangle_{\rho}^{ \rho_{a_i}\rho_{\overline{b}^{(i)}_2}}  \langle \pi_{\overline{a}_i}\pi_{\overline{b}^{(i)}_2}\rangle\langle \pi_{a_i}\pi_{\overline{b}^{(i)}_2}\rangle \Big\}.
\end{eqnarray}
Eq. (\ref{nl2}) cannot  acquire a negative value  if {\it none} of the twelve weak values in (\ref{nl2weak}) is anomalous. If a set of six weak values, $\Big(\llangle \pi_{b^{(i)}_1}\rrangle_{\rho}^{\rho_{a_i}\rho_{b^{(i)}_2}}$  and $\llangle  \pi_{\overline{b}^{(i)}_1}\rrangle_{\rho}^{\rho_{\overline{a}_i}\rho_{\overline{b}^{(i)}_2}}\Big)$, or $\Big(\llangle \pi_{b^{(i)}_1}\rrangle_{\rho}^{\rho_{\overline{a}_i}\rho_{b^{(i)}_2}}$ and $\llangle \pi_{\overline{b}^{(i)}_1}\rrangle_{\rho}^{ \rho_{a_i}\rho_{\overline{b}^{(i)}_2}}\Big)$, are anomalous, the ensuing inequality (\ref{E_nl_2}) is bound to get satisfied,
which again shows the intimate connection between anomalous weak values and entanglement.
%----------------------------------------------
 \subsubsection*{ Inequality III}
The last inequality, which we derive below, differs from the previous ones in that it  has contributions from both  local observables and correlations. 
     
The combination (which involves terms bilinear in pseudo-probabilities) is chosen to be,
 \begin{eqnarray}
         \label{nl3}
   {\cal S}_3  & = &  \sum_{i=1}^3  \Big[\mathcal{P}(a_i = b^{(i)}_1 = b^{(i)}_2)
     +\dfrac{1}{2}\Big\{ P(a_i)\mathcal{P}(\overline{a}^{(i)}_1, \overline{a}^{(i)}_2) \nonumber \\
&+ & P(\overline{a}_i)\mathcal{P}(a^{(i)}_1, a^{(i)}_2) 
+ P(b_i)\mathcal{P}(\overline{b}^{(i)}_1, \overline{b}^{(i)}_2) \nonumber \\
&+&  P(\overline{b}_i)\mathcal{P}(b^{(i)}_1, b^{(i)}_2) 
      +  P(a_i)\mathcal{P}(\overline{b}^{(i)}_1, \overline{b}^{(i)}_2)  \nonumber \\
&+  & P(\overline{a}_i)\mathcal{P}(b^{(i)}_1, b^{(i)}_2) 
     +  \mathcal{P}(\overline{a}^{(i)}_1, \overline{a}^{(i)}_2) P(b_i) \nonumber \\
&+  & \mathcal{P}(a^{(i)}_1, a^{(i)}_2) P(\overline{b}_i)\Big\}\Big].
   \end{eqnarray}

   The ensuing expression has the form ($\lambda= \frac{1}{2}\cos\frac{\alpha}{2}$),
 \begin{align}
   \label{Ent_nl_3}
  {\cal S}_3 =   \lambda\Big\{9 \cos \frac{\alpha}{2} &+ \sum_{i=1}^3\Big(\big\langle\vec{\sigma}_1\cdot\hat{a}_i\vec{\sigma}_2\cdot\hat{b}_i 
 \big\rangle-\frac{1}{2} \langle\vec{\sigma}_1\cdot\hat{a}_i + \vec{\sigma}_2\cdot\hat{b}_i\rangle^2\Big) \Big\}.
   \end{align}
   The non-classicality  condition,  ${\cal S}_3 < 0$, is enforced on the states with the proviso that,  ${\cal S}_3 \ge 0$, for all the separable states. This  determines the range of the free parameter  to be $0 < \alpha \le  \arccos(-79/81)$. The inequality corresponding to the upper limit, $\alpha = \arccos\Big(-\frac{79}{81}\Big)$, has been obtained by G\"uhne by using covariance matrix criteria for local observables \cite{Guhne04}.
   
   Writing equation (\ref{nl3}) in terms of weak values, the following expression results,
   \begin{eqnarray}
 {\cal S}_3 & = &    \sum_{i=1}^3\Big[\Big\{ \langle \pi_{a_i}\pi_{b^{(i)}_2}\rangle \llangle \pi_{b^{(i)}_1}\rrangle_{\rho}^{ \pi_{a_i}\pi_{b^{(i)}_2}} + \langle \pi_{\overline{a}_i}\pi_{\overline{b}^{(i)}_2}\rangle \llangle \pi_{\overline{b}^{(i)}_1}\rrangle_{\rho}^{ \pi_{\overline{a}_i}\pi_{\overline{b}^{(i)}_2}} \Big\}\nonumber\\
& + & \dfrac{1}{2}\Big\{ \langle\pi_{a_i}\rangle\langle\pi_{\overline{a}^{(i)}_1}\rangle \llangle \pi_{\overline{a}^{(i)}_2}\rrangle_{\rho}^{ \rho_{\overline{a}^{(i)}_1}} +\langle\pi_{\overline{a}_i}\rangle\langle\pi_{a^{(i)}_1}\rangle\llangle \pi_{a^{(i)}_2}\big\rrangle_{\rho}^{ \rho_{a^{(i)}_1}}+  \langle\pi_{b_i}\rangle\langle\pi_{\overline{b}^{(i)}_1}\rangle \llangle \pi_{\overline{b}^{(i)}_2}\big\rrangle_{\rho}^{ \rho_{\overline{b}^{(i)}_1}} +\langle\pi_{\overline{b}_i}\rangle\langle\pi_{b^{(i)}_1}\rangle\llangle \pi_{b^{(i)}_2}\rrangle_{\rho}^{ \rho_{b^{(i)}_1}}  \nonumber\\
&+& \langle\pi_{a_i}\rangle\langle\pi_{\overline{b}^{(i)}_1}\rangle \llangle \pi_{\overline{b}^{(i)}_2}\rrangle_{\rho}^{ \rho_{\overline{b}^{(i)}_1}}+ \langle\pi_{\overline{a}_i}\rangle\langle\pi_{b^{(i)}_1}\rangle \llangle \pi_{b^{(i)}_2}\rrangle_{\rho}^{ \rho_{b^{(i)}_1}}  +  \langle\pi_{b_i}\rangle \langle\pi_{\overline{a}^{(i)}_1}\rangle\llangle \pi_{\overline{a}^{(i)}_2}\rrangle_{\rho}^{ \rho_{\overline{a}^{(i)}_1}}+
\langle\pi_{\overline{b}_i}\big\rangle\langle\pi_{a^{(i)}_1}\rangle \llangle \pi_{a^{(i)}_2}\rrangle_{\rho}^{ \rho_{a^{(i)}_1}}\Big\}\Big].\nonumber\\
\label{nl3_weak}
   \end{eqnarray}
  The detailed derivation is given in appendix (\ref{Entnlinapp_3}).
  
      At this juncture, we note that there does exist partial positive transpose (PPT) criterion \cite{Peres96}, which identifies all the entangled states in $2 \otimes 2$ systems. In this paper, our aim is to systematically study violations of classical probability rules to obtain non-classicality conditions (not restricted only to entanglement), and their experimental implementation through weak measurements. What we have shown here is that the derived conditions also serve to detect entanglement, contingent on appropriate choices of observables. 

  \subsection{Two-qubit quantum discord  and weak values}
      \label{Discord}
A system displays non-classicality even if any  one weak value turns out to be anomalous. This statement is equivalent to recognising non classicality if even one pseudo-probability becomes negative. We may, therefore, relax  the condition that, $\mathcal{P}_{E} \ge 0$, for separable states,  and ask if  conditions for quantum discord \cite{Ollivier01} can be derived.  
We show below that the answer is in the affirmative.  The proof is by explicit construction. 
      
  By definition, a state whose discord, ${\cal D}^{1 \rightarrow 2}$ vanishes has the structure,
      \begin{eqnarray}
      \label{non-discordant}
      \rho^{12} = \sum_k p_k|\phi^1_{k}\rangle\langle \phi^1_{k}|\otimes \rho^2_k,
      \end{eqnarray}
      where, $\sum_k p_k|\phi^1_{k}\rangle\langle \phi^1_{k}\vert$ is the resolution of $\rho^1$
  in its eigenbasis.
 %the states  $\{|\phi^A_k\rangle\}$ are orthonormal and $p_k\geq 0, \sum_k p_k =1$. 

\begin{figure}[!htb]
\includegraphics[width = 0.5\textwidth]{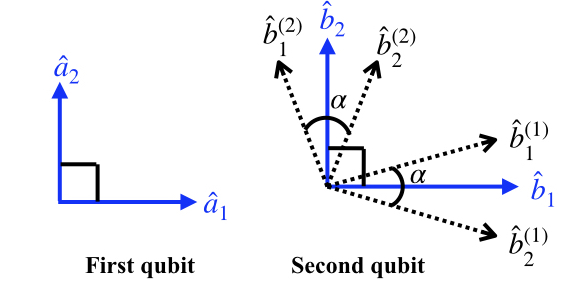}
\centering
\caption{\label{discord} $\hat{a}_i, \hat{b}_i$: Directions for first and second qubit  respectively that appear in the condition for discord (shown in blue). \\$\hat{b}_i^{(1)}, \hat{b}_i^{(2)}, i \in \{1 ,2\}:$ Directions for second qubit involved in the construction of pseudo-projections (shown in black).}
\end{figure}

\noindent {\bf Choice of the observables}: We choose two orthogonal observables $A_1, A_2$ for the first subsystem 1 with the stipulation that  one of them say, $A_1$ shares its eigenbasis with $\rho^1$.  The eigen-basis of $A_2$ is unbiased with respect to that of $A_1$. Thus, a partial tomography is warranted. For  this reason, the condition that we derive does not qualify as   a witness. For the subsystem 2,  we choose  two pairs of observables $\{B_1^{(1)}, B_2^{(1)}\} $, and $\{B_1^{(2)}, B_2^{(2)}\} $. They obey  the same conditions that were imposed in entanglement inequality.
The geometry is shown in Fig.(\ref{discord}).

Unlike in the earlier cases,  we construct   two pseudo-probabilities, ${\cal P}_D^i,  ~ i=1,2$, 
\begin{eqnarray}
\label{Cond}
 {\cal P}_D ^{i} &= & {\cal P}(a_i= b^{(i)}_1=b^{(i)}_2)  + 
  P(a_i) {\cal P}(\overline{b}^{(i)}_1\overline{b}^{(i)}_2) 
+ P(\overline{a}_i) {\cal P}(b^{(i)}_1 b^{(i)}_2) \nonumber \\
& \equiv & \lambda\Big\{(16\lambda + 2\langle  \vec{\sigma}_{1}\cdot\hat{a}_i\vec{\sigma}_{2}\cdot\hat{b}_i\rangle 
 - 2   \langle  \vec{\sigma}_{1}\cdot\hat{a}_i\rangle\langle \vec{\sigma}_{2}\cdot\hat{b}_i\rangle\Big\},
 \end{eqnarray}
where the parameter $\lambda = \dfrac{1}{4}\cos\dfrac{\alpha}{2}$.  If a state is discordant,
both the pseudo-probabilities  --   ${\cal P}^1_D$ and ${\cal P}^2_D$,  become  negative for  suitable choices of $\alpha$. The non-discordant states  can have  a negative value for at most one pseudo-probability. The detailed derivation is given in appendix (\ref{Discord_app}).

Weak measurements can indeed detect discord. Abstracting the parent PP from the LHS of Eq (\ref{Cond}), we infer the  expression, in terms of weak measurements,  to be,
\begin{eqnarray}
 {\cal P}^i_D & =  & \langle \pi_{a_i}\pi_{b^{(i)}_2}\rangle \llangle \pi_{b^{(i)}_1}\rrangle_{\rho}^{ \rho_{a_i}\rho_{b^{(i)}_2}}+ \langle \pi_{\overline{a}_i}\pi_{\overline{b}^{(i)}_2}\rangle \llangle \pi_{\overline{b}^{(i)}_1}\rrangle_{\rho}^{\rho_{\overline{a}_i}\rho_{\overline{b}^{(i)}_2}} \nonumber \\
& + & \langle \pi_{a_i}\rangle\langle \pi_{\overline{b}^{(i)}_2}\rangle \llangle \pi_{\overline{b}^{(i)}_1}\rrangle_{\rho}^{\rho_{\overline{b}^{(i)}_2}}+\langle \pi_{\overline{a}_i}\rangle\langle \pi_{b^{(i)}_2}\rangle \llangle \pi_{b^{(i)}_1}\rrangle_{\rho}^{\rho_{b^{(i)}_2}}.
\end{eqnarray}
 It is clear that the existence of quantum discord is concomitant on the existence of negative weak values.  

We conclude this section with a final comment. Through these constructions, we have explicitly shown that non-existence of classical joint probability underlies quantum entanglement and discord as well, a conclusion similar to that obtained by Fine \cite{Fine82} for CHSH nonlocality. Weak measurement techniques provide an experimental tool to bring this aspect out through anomalous weak values.
%---------------------------------------------
\section{Discussion on the choices of pseudoprobabilities for different quantum features}
\label{Discussion_choice}
We are now in a position to discuss how the hierarchy among nonlocality, entanglement, discord and coherence gets manifested in the choices of pseudoprobabilities. The hierarchy is manifested in two ways-- number of observables used in the pseudo-probability scheme and the way they are combined.

Consider, for example, two-qubit states, in which CHSH nonlocality is the strongest kind of nonclassicality. Thus, only four observables are needed to arrive at CHSH inequality, which is the minimal requirement.

Entanglement is weaker than nonlocality, but stronger than quantum discord. Recall that there are local entangled  states and separable discordant states. Consequently, entanglement inequalities emerge from the pseudoprobability schemes involving larger number of observables. For example, the pseudo-probability scheme underlying the entanglement inequality (\ref{Ent2qubit}) is constructed for joint outcomes of six observables (two observables for the first qubit and four observables for the second qubit). On the other hand, the pseudo-probability scheme, underlying the entanglement inequality (\ref{Ent2qubit1}), requires joint outcomes of nine observables (three observables for the first qubit and six for the second).  Derivation of even more stringent nonlinear entanglement inequalities would require more involved sums of pseudo-probabilities.

Finally, detection of quantum discord would also require partial tomography. Thus, the choices of pseudoprobabilities employed by us are not haphazard, but reflect the strength of underlying nonclassical feature.
Coherence, of course, is the simplest as it does not involve correlations.

%----------------------------------------------------------------
 %-------------------------------------------------------
 \section{Conclusion}
 \label{Conclusion}
  In conclusion, we have established the connection between two seemingly different approaches to non-classicality by making use of pseudo-projections \cite{Adhikary18} --  in fact a special case of which leads to the non classical properties described in \cite{Hill61, Barut88}. Pseudo-probabilities  and anomalous weak values share an equivalence which makes the former measurable quantities. Violation of Boolean logic by quantum mechanics can be directly probed by employing weak measurements.  We have shown how different sets of anomalous weak values not only serve to detect different nonclassicality manifests in single and two- qubit states but also trace the violations of underlying classical probability rules.  
  
   A further extension of our results to multi-party systems constitutes an interesting study. It will integrate conditions for different kinds of nonclassical correlations (e.g., nonlocality, entanglement etc.), anomalous weak values and negative pseudoprobabilities.

 %-------------------------------------------------------
 \section*{Acknowledgement}
It is a pleasure to thank Lev Vaidman for illuminating discussions  and bringing several relevant works to our notice, and  Fabrizio Piacentini for a fruitful  discussion.  We would like to thank the anonymous referees whose comments have helped to enhance the quality of the manuscript. We thank Rajni Bala for several discussions and helpful suggestions. Sooryansh  thanks the Council for Scientific and Industrial Research (Grant no. -09/086 (1278)/2017-EMR-I) for funding his research.

\begin{appendices}
\section{Unit pseudo-projections have at least one negative eigenvalue.}
\label{Eigen_conjunction}
Consider a unit PP, ${\bf \Pi}$, constructed as the symmetrised product of $N$ mutually non-commuting projections (of any ranks) in  a Hilbert space of dimension $D$,   as given by,

\begin{equation}
{\bf \Pi} = \dfrac{1}{2}(\pi_1 \pi_2\cdots\pi_N + \rm{h. c. }).
\end{equation}

Let $\vert a \rangle~ (\vert b \rangle)$ be a  vector in the null space of $\pi_1 ~(\pi_N)$ but not in that of $\pi_N ~(\pi_1)$.  Construct the orthogonal basis: $\{\vert e_1\rangle =\vert a \rangle$, $\vert e_2\rangle = \vert b\rangle -\langle a \vert b \rangle \vert a\rangle\}$ in the two dimensional subspace spanned by the vectors. The determinant of the  
 minor of ${\bf \Pi}$ in this two dimensional basis is evidently  negative. Thus, ${\bf \Pi}$  possesses atleast  one negative eigenvalue.
In order to illustrate it, consider a unit PP, ${\bf \Pi}$, constructed as symmetrised product of $N$ mutually non-commuting projections $\pi_j = \sum_{i_j}|a_{i_j}\rangle\langle a_{i_j}|;~j \in \{1, \cdots, N\}$. Thus,
\begin{align}
{\bf \Pi} = \frac{1}{2}\sum_{i_1 \cdots i_N} \Big(|a^1_{i_1}\rangle\langle a^1_{i_1}|a^2_{i_2}\rangle\langle a^2_{i_2}|\cdots\langle a^{N-1}_{i_{N-1}}|a^N_{i_N}\rangle\langle a^N_{i_N}| +{\rm h.c.} \Big).\nonumber
\end{align}
Let $i_1 \in \{1, \cdots, d\}$, where $d < D$. Since $\pi_1$ and $\pi_N$ are non-commuting, one can always find a state $|a^1_{d+1}\rangle$, lying in the null space of $\pi_1$, such that $\pi_N|a^1_{d+1}\rangle \neq 0$. We project ${\bf \Pi}$ in the subspace spanned by $\{|a^1_d\rangle, |a^1_{d+1}\rangle\}$,
\begin{align}
& \Big(|a^1_d\rangle\langle a^1_d| +|a^1_{d+1}\rangle\langle a^1_{d+1}|\Big){\bf \Pi}\Big(|a^1_d\rangle\langle a^1_d| +|a^1_{d+1}\rangle\langle a^1_{d+1}|\Big)\nonumber\\
=&\frac{1}{2}\sum_{i_1\cdots i_N}\Big\{|a^1_d\rangle\langle a^1_{i_1}|a^2_{i_2}\rangle\langle a^2_{i_2}|\cdots\langle a^{N-1}_{i_{N-1}}|a^N_{i_N}\rangle\Big(\langle|a^N_{i_N}|a^1_d\rangle \langle a^1_d| +\langle a^N_{i_N}|a^1_{d+1}\rangle\langle a^1_{d+1}|\Big) + {\rm h.c.}\Big\}\nonumber\\
=&\frac{1}{2}|a^1_d\rangle\Big\{\sum_{i_1\cdots i_N}\langle a^1_{i_1}|a^2_{i_2}\rangle\langle a^2_{i_2}|\cdots\langle a^{N-1}_{i_{N-1}}|a^N_{i_N}\rangle\langle a^N_{i_N}|a^1_d\rangle + {\rm c.c.}\Big\}\langle a^1_d|\nonumber\\
=&\frac{1}{2}|a^1_d\rangle \Big\{\sum_{i_1\cdots i_N}\langle a^1_{i_1}|a^2_{i_2}\rangle \cdots\langle a^N_{i_{N-1}}|a^N_{i_N}\rangle\langle a^N_{i_N}|a^1_{d+1}\rangle\Big\}\langle a^1_{d+1}| 
+{\rm h.c.},
\end{align}
which has a negative determinant, thereby proving that ${\bf \Pi}$ has a negative eigenvalue.

\section{Derivation of coherence witness}
\label{appcoherence}
 In a fixed basis, all the diagonal states are considered to be incoherent and states having non-zero off-diagonal elements are designated as  coherent \cite{Streltsov17}. PPs can act as witnesses for this coherence. To see this, let the state be diagonal in the eigenbasis of $\sigma_z$.
We are interested in the PP, representing  a joint outcome of two incompatible observables $\vec{\sigma}\cdot\hat{a}_1$ and $\vec{\sigma}\cdot\hat{a}_2$, which acquires  negative values only for states $\rho = \frac{1}{2} ({1} + \vec{\sigma}.\vec{p})$ having non-zero off-diagonal term. 
Then, the PP, 
${\bf \Pi}_{a_1a_2} = \frac{1}{4} \Big({1} + \hat{a}_1\cdot\hat{a}_2 + \vec{\sigma}.(\hat{a}_1 + \hat{a}_2)\Big)$, 
has positive overlap with all diagonal states whenever $(\hat{a}_1 +\hat{a}_2)$ lies in the $x-y$ plane. 
Suppose $\hat{a}_1\cdot\hat{a}_2 = \cos \theta$ and we choose $(\hat{a}_1 + \hat{a}_2) \parallel (\cos \lambda \hat{x} + \sin \lambda \hat{y})$. Then,
\begin{eqnarray}
{\bf \Pi}_{a_1a_2} = \frac{1}{2}\cos \frac{\theta}{2}\Big\{ \cos \frac{\theta}{2} + (\cos \lambda \sigma_x + \sin \lambda \sigma_y)\Big\}.
\end{eqnarray}
The expectation value of this operator with the state $\rho$ is,
\begin{eqnarray}
{\rm Tr}({\bf \Pi}_{a_1a_2}  \rho) = \frac{1}{2}\cos \frac{\theta}{2} \Big\{\cos \frac{\theta}{2} + (\cos \lambda p_x + \sin \lambda p_y)\Big\}.
\end{eqnarray}
The operator ${\bf \Pi}_{a_1a_2}$, for a proper choice of $\theta$ and $\lambda$, has negative overlap for all states with non-zero $p_x$ and $p_y$.

\section{Derivation of CHSH inequality}
\label{Belldetailed}
The sum of pseudoprobabilities for CHSH nonlcoality is as follows:
\begin{align}
&{\cal P}(A_1=B_1=B_2)+{\cal P}(A_2=B_1={\bar B}_2).
\end{align}
Employing dichotomic nature of the observables, we  express each pseudoprobability in terms of expectation values of observables. Thus, the following expression results,
\begin{align}
{\cal P}_{\rm NL} \equiv&{\cal P}(A_1=B_1=B_2)+{\cal P}(A_2=B_1={\bar B}_2)\nonumber\\
=&{\cal P}(A_1=+1;B_1=+1B_2=+1)+ {\cal P}(A_1=-1;B_1=-1B_2=-1)\nonumber\\
+&{\cal P}(A_2=+1;B_1=+1{B}_2=-1)+{\cal P}(A_2=-1;B_1=-1{B}_2=+1)\nonumber\\
=&\dfrac{1}{8}\Big\langle\big(1+A_1\big)\big(1+\{B_1, B_2\}+B_1+B_2\big)+\big(1-A_1\big)\big(1+\{B_1, B_2\}-B_1-B_2\big)\nonumber\\
+&\big(1+A_2\big)\big(1-\{B_1, B_2\}+B_1-B_2\big)+\big(1-A_2\big)\big(1-\{B_1, B_2\}-B_1+B_2\big)\Big\rangle\nonumber\\
=&\dfrac{1}{4}\Big\langle 2+A_1(B_1+B_2)+A_2(B_1-B_2)\Big\rangle.
\end{align}
Imposing the nonclassicality criteria, ${\cal P}_{\rm NL}<0$, the CHSH inequality results,
\begin{align}
\langle A_1(B_1+B_2)+A_2(B_1-B_2)\rangle <-2.
\end{align}
For the following choice of the observables, 
\begin{align}
& A_1\equiv \vec{\sigma}_1\cdot\hat{a}_1=\sigma_{1x}; A_2\equiv \vec{\sigma}_1\cdot\hat{a}_2=\sigma_{1y};\nonumber\\
& B_1\equiv \frac{\sigma_{2x}+\sigma_{2y}}{\sqrt{2}}; B_2\equiv \frac{\sigma_{2x}-\sigma_{2y}}{\sqrt{2}},
\end{align}
the values of the pseudo-probabilities, for two-qubit Werner states, $\rho = \frac{1}{4}(1-\eta\vec{\sigma}_1\cdot\vec{\sigma}_2)$, are given as below,
\begin{align}
& {\cal P}(A_1;B_1B_2) ={\cal P}({\bar A}_1;{\bar B}_1{\bar B}_2)= {\cal P}(A_2;B_1{\bar B}_2) ={\cal P}({\bar A}_1;{\bar B}_1B_2) =\dfrac{1}{8}(1-\eta\sqrt{2}).\nonumber
\end{align}
Thus, all the four pseudoprobabilities turn negative for all the nonlocal Werner states ($\frac{1}{\sqrt{2}}<\eta\leq 1$), thereby proving that all the weak values also turn negative, i.e., anomalous.
\section{Derivation of linear entanglement inequalities}
\subsection{Expression of pseudoprojection operator for joint outcomes of two observables}
First, we explicitly calculate the expression of PP representing a joint event, in which   $\vec{\sigma}\cdot\hat{m}_1$ and  $\vec{\sigma}\cdot\hat{m}_2$ both take value $+1$. The PP is given by the symmetrised rule,
\begin{align}
{\bf \Pi} &=\dfrac{1}{2}\Big\{\dfrac{1}{2}(1+\vec{\sigma}\cdot\hat{m}_1), \dfrac{1}{2}(1+\vec{\sigma}\cdot\hat{m}_2)\Big\}\nonumber\\
&=\dfrac{1}{4}\Big(1+\frac{1}{2}\{\vec{\sigma}\cdot\hat{m}_1, \vec{\sigma}\cdot\hat{m}_2\}+\vec{\sigma}\cdot(\hat{m}_1+\hat{m}_2)\Big)\nonumber\\
\label{PPC}
&=\dfrac{1}{4}\Big(1+\hat{m}_1\cdot\hat{m}_2+\vec{\sigma}\cdot\hat{m}_1+\vec{\sigma}\cdot\hat{m}_2\Big).
\end{align}
If the included angle between $\hat{m}_1$ and $\hat{m}_2$ is given by $\alpha$, then,
$$\hat{m}_1\cdot\hat{m}_2 = \cos\alpha~ {\rm and}~ \hat{m}+\hat{m}_2 = 2\cos\frac{\alpha}{2}\hat{m}.$$ Here, $\hat{m}$ is a unit vector parallel to $(\hat{m}_1+\hat{m}_2)$. Plugging in these values in equation (\ref{PPC}), we obtain,
\begin{align}
{\bf \Pi} &= \dfrac{1}{4}\Big(2\cos^2\frac{\alpha}{2}+2\cos\frac{\alpha}{2}\vec{\sigma}\cdot\hat{m}\Big)\nonumber\\
\label{PPexpression}
&= \dfrac{1}{2}\cos\frac{\alpha}{2}\Big(\cos\frac{\alpha}{2}+\vec{\sigma}\cdot\hat{m}\Big).
\end{align}
We shall thoroughly use this expression to obtain various entanglement inequalities, conditions for discord and coherence witnesses.We start with the derivation of the first linear entanglement inequality. 
\subsection{Derivation of linear entanglement inequality I}
\label{Entlinapp_1}
The sum of pseudoprobabilities for entanglement inequality is as follows:
\begin{align}
{\cal P}_{E_1} = \sum_{i=1}^2{\cal P}(a_i=b^{(i)}_1=b^{(i)}_2).\nonumber
\end{align}
Writing each pseudo-probability in terms of expectation of PP, the following expression results,
\begin{align}
& {\cal P}(a_i = b_1^{(i)}=b_2^{(i)})={\cal P}(a_i = +1;b_1^{(i)}=+1b_2^{(i)}=+1)+{\cal P}(a_i =-1; b_1^{(i)}=-1b_2^{(i)}=-1)\nonumber\\
=&\dfrac{1}{4}{\cos\frac{\alpha}{2}}\Big\langle(1+\vec{\sigma}_1\cdot\hat{a}_i)(\cos\frac{\alpha}{2}+\vec{\sigma}_2\cdot\hat{b}_i)+(1-\vec{\sigma}_1\cdot\hat{a}_i)(\cos\frac{\alpha}{2}-\vec{\sigma}_2\cdot\hat{b}_i)\Big\rangle\nonumber\\
=&\dfrac{1}{2}{\cos\frac{\alpha}{2}}\Big\langle\cos\dfrac{\alpha}{2}+\vec{\sigma}_1\cdot\hat{a}_i\vec{\sigma}_2\cdot\hat{b}_i\Big\rangle.
\end{align}
Substituting the values of pseudoprobabilities in terms of expectation values of observables, we obtain the following expression,
\begin{align}
{\cal P}_{E_1} = \dfrac{1}{2}{\cos\frac{\alpha}{2}}\sum_{i=1}^2\big\langle 2\cos\frac{\alpha}{2}+\vec{\sigma}_1\cdot\hat{a}_i\vec{\sigma}_2\cdot\hat{b}_i\big\rangle.
\end{align} 
Imposing the nonclassicality constraint, ${\cal P}_{E_1} <0$, the following inequality emerges,
\begin{align}
\label{App:inequ1}
\sum_{i=1}^2\big\langle 2\cos\frac{\alpha}{2}+\vec{\sigma}_1\cdot\hat{a}_i\vec{\sigma}_2\cdot\hat{b}_i\big\rangle<0.
\end{align}
The range of $\alpha$ is fixed since we demand all the separable states to violate the inequality. The maximum value of $\sum_{i=1}^2\langle\vec{\sigma}_1\cdot\hat{a}_i\vec{\sigma}_2\cdot\hat{b}_i \rangle $ for a separable state is 1. To show this, consider the two--qubit pure separable state $\rho = \dfrac{1}{4}(1+\sigma_{1z})(1+\sigma_{2z})$. Then,
\begin{align}
& \sum_{i=1}^2\big\langle 2\cos\frac{\alpha}{2}+\vec{\sigma}_1\cdot\hat{a}_i\vec{\sigma}_2\cdot\hat{b}_i\big\rangle_{\rho}\nonumber\\
=& 2\cos\frac{\alpha}{2}+a_{1z}b_{1z}+a_{2z}b_{2z} \geq -1+2\cos\frac{\alpha}{2}.
\end{align}
Note that, $\hat{a}_1\perp\hat{a}_2$ and $\hat{b}_1\perp\hat{b}_2$. It implies that $(a_{1z}b_{1z}+a_{2z}b_{2z})$ is upper bounded by $1$ in magnitude.
Thus, in order that the inequality (\ref{App:inequ1}) gets violated by all the separable states, $1 \leq 2\cos\frac{\alpha}{2}< 2$, which, in turn, implies that $ 0\leq \alpha <\frac{2\pi}{3}$.

In terms of weak values, the sum of pseudoprobabilities, ${\cal P}_{E_1}$, can be written as,
\begin{align}
{\cal P}_{E_1} &= \sum_{i=1}^2{\cal P}(a_i=b^{(i)}_1=b^{(i)}_2)\nonumber\\
&=\sum_{i=1}^2\big\{{\cal P}(a_i=+1;b^{(i)}_1=+1b^{(i)}_2=+1)+{\cal P}(a_i=-1;b^{(i)}_1=-1b^{(i)}_2=-1)\big\}\nonumber\\
&=\sum_{i=1}^2\langle \pi_{a_i}\otimes{\bf \Pi}_{b^{(i)}_1b^{(i)}_2} + \pi_{{\bar a}_i}\otimes{\bf \Pi}_{{\bar b}^{(i)}_1{\bar b}^{(i)}_2}\rangle\nonumber\\
&=\dfrac{1}{2}\sum_{i=1}^2\langle \pi_{a_i}\otimes\{{\pi}_{b^{(i)}_1},\pi_{b^{(i)}_2}\} + \pi_{{\bar a}_i}\otimes\{{\pi}_{{\bar b}^{(i)}_1}, \pi_{{\bar b}^{(i)}_2}\}\rangle\nonumber\\
&=\sum_{i=1}^2\Big(\langle \pi_{a_i}\pi_{b^{(i)}_1}\rangle\llangle\pi_{b^{(i)}_2}\rrangle_{\rho}^{\rho_{a_i}\rho_{b^{(i)}_1}} + \langle \pi_{{\bar a}_i}\pi_{{\bar b}^{(i)}_1}\rangle\llangle\pi_{{\bar b}^{(i)}_2}\rrangle_{\rho}^{\rho_{{\bar a}_i}\rho_{{\bar b}^{(i)}_1}} \Big).
\end{align}
For the special choice of observables,
\begin{align}
\vec{\sigma}_1\cdot\hat{a}_1=\sigma_{1x}; \vec{\sigma}_1\cdot\hat{a}_2=\sigma_{1y}; \vec{\sigma}_2\cdot\hat{b}_1=\sigma_{2x};  \vec{\sigma}_2\cdot\hat{b}_2=\sigma_{2y},
\end{align}
and for the $2\otimes 2$ werner states, $\rho = \dfrac{1}{4}(1-\eta\vec{\sigma}_1\cdot\vec{\sigma}_2)$, the pseudoprobabilities and the inequality assumes the following form,
\begin{align}
&{\cal P}(a_1 = +1;b_1^{(1)}=+1b_2^{(1)}=+1)=
{\cal P}(a_1 =-1; b_1^{(1)}=-1b_2^{(1)}=-1)  \nonumber\\
=&{\cal P}(a_2 = +1;b_1^{(2)}=+1b_2^{(2)}=+1)= {\cal P}(a_2 =-1; b_1^{(2)}=-1b_2^{(2)}=-1) \nonumber\\
=&\dfrac{1}{4}\cos\dfrac{\alpha}{2} \Big(\cos\dfrac{\alpha}{2}-\eta\Big) \nonumber\\
&2\cos\dfrac{\alpha}{2}+\langle\sigma_{1x}\sigma_{2x}+\sigma_{1y}\sigma_{2y}\rangle<0.
\end{align}
We consider a particular value of $\alpha = \dfrac{2\pi}{3}$.  For this value, all the four pseudoprobabilities are equal to the following value,
\begin{align}
& {\cal P}(a_1 = +1;b_1^{(1)}=+1b_2^{(1)}=+1)= {\cal P}(a_1 =-1; b_1^{(1)}=-1b_2^{(1)}=-1)\nonumber\\
=& {\cal P}(a_2 = +1;b_1^{(2)}=+1b_2^{(2)}=+1)= {\cal P}(a_2 =-1; b_1^{(2)}=-1b_2^{(2)}=-1)\nonumber\\
=&\dfrac{1}{8}\Big(\dfrac{1}{2}-\eta\Big).
\end{align} 
The Werner states are detected to be entangled in the range $\eta \in \Big(\dfrac{1}{2}, 1\Big]$ by the inequality (\ref{Ent2qubit}), which implies that all the four pseudoprobabilities will be negative. This, in turn, implies that all the four weak values are negative, i.e., anomalous.

\subsection{Derivation of linear entanglement inequality: II}
\label{Entlinapp_2}
The sum of pseudoprobabilities for entanglement inequality is as follows:
\begin{align}
{\cal P}_{E_2} = \sum_{i=1}^3{\cal P}(a_i=b^{(i)}_1=b^{(i)}_2).\nonumber
\end{align}
The pseudoprobabilities, as before, can be written as expectation values of Pauli operators as follows,
\begin{align}
{\cal P}(a_i = b_1^{(i)}=b_2^{(i)})=&\dfrac{1}{4}\cos\frac{\alpha}{2}\Big\langle(1+\vec{\sigma}_1\cdot\hat{a}_i)(\cos\frac{\alpha}{2}+\vec{\sigma}_2\cdot\hat{b}_i)
+(1-\vec{\sigma}_1\cdot\hat{a}_i)(\cos\frac{\alpha}{2}-\vec{\sigma}_2\cdot\hat{b}_i)\Big\rangle\nonumber\\
=&\dfrac{1}{2}{\cos\frac{\alpha}{2}}\Big\langle\cos\dfrac{\alpha}{2}+\vec{\sigma}_1\cdot\hat{a}_i\vec{\sigma}_2\cdot\hat{b}_i\Big\rangle.
\end{align}
Substituting the values of pseudoprobabilities in terms of expectation values of observables, we obtain the following expression,
\begin{align}
{\cal P}_{E_2} =\dfrac{1}{2} {\cos\frac{\alpha}{2}}\sum_{i=1}^3\big\langle 2\cos\frac{\alpha}{2}+\vec{\sigma}_1\cdot\hat{a}_i\vec{\sigma}_2\cdot\hat{b}_i\big\rangle.
\end{align} 
The nonclassicality condition,   ${\cal P}_{E_2}< 0$, yields the following inequality,
 \begin{align}
 \label{App:ineq2}
\sum_{i=1}^3\big\langle 2\cos\frac{\alpha}{2}+\vec{\sigma}_1\cdot\hat{a}_i\vec{\sigma}_2\cdot\hat{b}_i\big\rangle <0.
\end{align} 
The range of $\alpha$ is fixed since we demand all the separable states to violate the inequality (\ref{App:ineq2}). The maximum value of $\sum_{i=1}^3\langle\vec{\sigma}_1\cdot\hat{a}_i\vec{\sigma}_2\cdot\hat{b}_i \rangle $ for a separable state is 1. To show this, consider the two--qubit pure separable state $\rho = \dfrac{1}{4}(1+\sigma_{1z})(1+\sigma_{2z})$. Then,
\begin{align}
& \sum_{i=1}^3\big\langle 3\cos\frac{\alpha}{2}+\vec{\sigma}_1\cdot\hat{a}_i\vec{\sigma}_2\cdot\hat{b}_i\big\rangle_{\rho}\nonumber\\
=& 3\cos\frac{\alpha}{2}+a_{1z}b_{1z}+a_{2z}b_{2z}+a_{3z}b_{3z} \geq -1+3\cos\frac{\alpha}{2}.
\end{align}
Thus, in order that the inequality gets violated by all the separable states, $1 \leq 3\cos\frac{\alpha}{2}< 3$, which, in turn, implies that $ 0\leq \alpha <{\rm arccos}\big(-\frac{7}{9}\big)$.

In terms of weak values, the sum of pseudoprobabilities ${\cal P}_{E_2}$ can be written as,
\begin{align}
{\cal P}_{E_2} &= \sum_{i=1}^3{\cal P}(a_i=b^{(i)}_1=b^{(i)}_2)\nonumber\\
&=\sum_{i=1}^3\big\{{\cal P}(a_i=+1;b^{(i)}_1=+1b^{(i)}_2=+1)+{\cal P}(a_i=-1;b^{(i)}_1=-1b^{(i)}_2=-1)\big\}\nonumber\\
&=\sum_{i=1}^3\langle \pi_{a_i}\otimes{\bf \Pi}_{b^{(i)}_1b^{(i)}_2} + \pi_{{\bar a}_i}\otimes{\bf \Pi}_{{\bar b}^{(i)}_1{\bar b}^{(i)}_2}\rangle\nonumber\\
&=\dfrac{1}{2}\sum_{i=1}^3\langle \pi_{a_i}\otimes\{{\pi}_{b^{(i)}_1},\pi_{b^{(i)}_2}\} + \pi_{{\bar a}_i}\otimes\{{\pi}_{{\bar b}^{(i)}_1}, \pi_{{\bar b}^{(i)}_2}\}\rangle\nonumber\\
&=\sum_{i=1}^3\Big(\langle \pi_{a_i}\pi_{b^{(i)}_1}\rangle\llangle\pi_{b^{(i)}_2}\rrangle_{\rho}^{\rho_{a_i}\rho_{b^{(i)}_1}} + \langle \pi_{{\bar a}_i}\pi_{{\bar b}^{(i)}_1}\rangle\llangle\pi_{{\bar b}^{(i)}_2}\rrangle_{\rho}^{\rho_{{\bar a}_i}\rho_{{\bar b}^{(i)}_1}} \Big).
\end{align}
For the special choices of observables,
\begin{align}
\vec{\sigma}_1\cdot\hat{a}_1=\sigma_{1x}; \vec{\sigma}_1\cdot\hat{a}_2=\sigma_{1y};  \vec{\sigma}_1\cdot\hat{a}_3=\sigma_{1z}; \vec{\sigma}_2\cdot\hat{b}_1=\sigma_{2x};  \vec{\sigma}_2\cdot\hat{b}_2=\sigma_{2y}, \vec{\sigma}_2\cdot\hat{b}_3=\sigma_{2z},
\end{align}
and for the $2\otimes 2$ werner states $\rho = \dfrac{1}{4}(1-\eta\vec{\sigma}_1\cdot\vec{\sigma}_2)$, the pseudoprobabilities and the inequality assumes the following form,
\begin{align}
&{\cal P}(a_1 = +1;b_1^{(1)}=+1b_2^{(1)}=+1)=
{\cal P}(a_1 =-1; b_1^{(1)}=-1b_2^{(1)}=-1)\nonumber\\
=&{\cal P}(a_2 = +1;b_1^{(2)}=+1b_2^{(2)}=+1)=
{\cal P}(a_2 =-1; b_1^{(2)}=-1b_2^{(2)}=-1)\nonumber\\
=&{\cal P}(a_3 = +1;b_1^{(3)}=+1b_2^{(3)}=+1)={\cal P}(a_3 =-1; b_1^{(3)}=-1b_2^{(3)}=-1)\nonumber\\
=&\dfrac{1}{4}\cos\dfrac{\alpha}{2} \Big(\cos\dfrac{\alpha}{2}-\eta\Big) \nonumber\\
&3\cos\dfrac{\alpha}{2}+\langle\sigma_{1x}\sigma_{2x}+\sigma_{1y}\sigma_{2y}+\sigma_{1z}\sigma_{2z}\rangle<0.
\end{align}
We consider the case $\alpha ={\rm arccos} \Big(-\dfrac{7}{9}\Big)$, then all the six pseudoprobabilities are equal to the following value,
\begin{align}
& {\cal P}(a_1 = +1;b_1^{(1)}=+1b_2^{(1)}=+1)= {\cal P}(a_1 =-1; b_1^{(1)}=-1b_2^{(1)}=-1)\nonumber\\
=& {\cal P}(a_2 = +1;b_1^{(2)}=+1b_2^{(2)}=+1)= {\cal P}(a_2 =-1; b_1^{(2)}=-1b_2^{(2)}=-1)\nonumber\\
=& {\cal P}(a_3 = +1;b_1^{(3)}=+1b_2^{(3)}=+1)= {\cal P}(a_3 =-1; b_1^{(3)}=-1b_2^{(3)}=-1)\nonumber\\
=&\dfrac{1}{8}\Big(\dfrac{1}{2}-\eta\Big).
\end{align} 
The Werner states are detected to be entangled in the range $\eta \in \Big(\dfrac{1}{3}, 1\Big]$ by the inequality (\ref{Ent2qubit1}), which implies that all the six pseudoprobabilities will be negative. This, in turn, implies that all the six weak values are negative, i.e., anomalous. 
\section{Derivation of non-linear entanglement inequalities}
\subsection{Derivation of non-linear entanglement inequality I}
\label{Entnlinapp_1}
The sum of pseudoprobabilities for entanglement inequality is as follows:
 \begin{equation}
 \label{nl11}
{\cal S}_1 =  \sum_{i=1}^2\mathcal{P}(a_i=b^{(i)}_1=b^{(i)}_2)\mathcal{P}(\overline{a}_i =b^{(i)}_1=b^{(i)}_2).
 \end{equation}
 As before, 
 \begin{align}
{\cal P}(a_i = b_1^{(i)}=b_2^{(i)})=&\dfrac{1}{4}{\cos\frac{\alpha}{2}}\Big\langle(1+\vec{\sigma}_1\cdot\hat{a}_i)(\cos\frac{\alpha}{2}+\vec{\sigma}_2\cdot\hat{b}_i)+(1-\vec{\sigma}_1\cdot\hat{a}_i)(\cos\frac{\alpha}{2}-\vec{\sigma}_2\cdot\hat{b}_i)\Big\rangle\nonumber\\
\label{NL1}
=&\dfrac{1}{2}{\cos\frac{\alpha}{2}}\Big\langle\cos\dfrac{\alpha}{2}+\vec{\sigma}_1\cdot\hat{a}_i\vec{\sigma}_2\cdot\hat{b}_i\Big\rangle,
\end{align}
 and,
  \begin{align}
{\cal P}({\bar a}_i = b_1^{(i)}=b_2^{(i)})=&\dfrac{1}{4}{\cos\frac{\alpha}{2}}\Big\langle(1-\vec{\sigma}_1\cdot\hat{a}_i)(\cos\frac{\alpha}{2}+\vec{\sigma}_2\cdot\hat{b}_i)\Big)+\dfrac{1}{4}\Big((1+\vec{\sigma}_1\cdot\hat{a}_i)(\cos\frac{\alpha}{2}-\vec{\sigma}_2\cdot\hat{b}_i)\Big\rangle\nonumber\\
\label{NL2}
=&\dfrac{1}{2}{\cos\frac{\alpha}{2}}\Big\langle\cos\dfrac{\alpha}{2}-\vec{\sigma}_1\cdot\hat{a}_i\vec{\sigma}_2\cdot\hat{b}_i\Big\rangle.
\end{align}
Plugging in the expressions from equations (\ref{NL1}) and (\ref{NL2}) in equation (\ref{nl11}),
\begin{align}
{\cal S}_1 =\Big(\dfrac{1}{2}{\cos\frac{\alpha}{2}}\Big)^2\sum_{i=1}^2\Big(2\cos^2\dfrac{\alpha}{2}-\langle\vec{\sigma}_1\cdot\hat{a}_i\vec{\sigma}_2\cdot\hat{b}_i\rangle^2\Big).
\end{align}
 Imposing the  non-classicality condition, ${\cal S}_1 < 0$,  we arrive at the inequality,
 \begin{equation}
 \label{Ineqnl}
 2\cos^2\dfrac{\alpha}{2} -\sum_{i=1}^2 \langle\vec{\sigma}_1\cdot\hat{a}_i\vec{\sigma}_2\cdot\hat{b}_i\rangle^2 <0.
 \end{equation}
 The range of $\alpha$ can be fixed by demanding all the separable states to violate this inequality. We can consider the separable state, without any loss of generality, to be $\rho_s=\dfrac{1}{4}(1+\sigma_{1z})(1+\sigma_{2z})$. For this state, the LHS of the inequality (\ref{Ineqnl}) assumes the following form,
  \begin{equation}
 2\cos^2\dfrac{\alpha}{2} - \langle a_{1z}b_{1z}\rangle^2-\langle a_{2z}b_{2z}\rangle^2\geq 2\cos^2\dfrac{\alpha}{2} - 1.
 \end{equation}
 The second step follows as, $\hat{a}_1\perp \hat{a}_2$ and $\hat{b}_1\perp\hat{b}_2$. In order that all the separable states violate this inequality, $ 1\leq  2\cos^2\dfrac{\alpha}{2}  < 2$, which, in turn, implies that, $0< \alpha \leq \frac{\pi}{2}$.
 In terms of weak values,
\begin{eqnarray}
{\cal S}_1 =  \sum_{i=1}^2\Big\{ &\llangle \pi_{b^{(i)}_1}\rrangle_{\rho}^{\rho_{a_i}\rho_{b^{(i)}_2}} \llangle \pi_{b^{(i)}_1}\rrangle_{\rho}^{\rho_{\overline{a}_i}\rho_{b^{(i)}_2}} \langle \pi_{a_i}\pi_{b^{(i)}_2}\rangle\langle \pi_{\overline{a}_i}\pi_{b^{(i)}_2}\rangle+\nonumber\\
& \llangle \pi_{\overline{b}^{(i)}_1}\rrangle_{\rho}^{\rho_{\overline{a}_i}\rho_{\overline{b}^{(i)}_2}}\llangle \pi_{\overline{b}^{(i)}_1}\rrangle_{\rho}^{ \rho_{a_i}\rho_{\overline{b}^{(i)}_2}}  \langle \pi_{\overline{a}_i}\pi_{\overline{b}^{(i)}_2}\rangle\langle \pi_{a_i}\pi_{\overline{b}^{(i)}_2}\rangle \Big\}.
\end{eqnarray}
 \subsection{Derivation of non-linear entanglement inequality: II}
 \label{Entnlinapp_2}
 The sum of pseudoprobabilities for entanglement inequality is as follows:
 \begin{equation}
{\cal S}_2 =  \sum_{i=1}^3\mathcal{P}(a_i=b^{(i)}_1=b^{(i)}_2)\mathcal{P}(\overline{a}_i =b^{(i)}_1=b^{(i)}_2).
 \end{equation}
 As before, 
 \begin{align}
{\cal P}(a_i = b_1^{(i)}=b_2^{(i)})=&\dfrac{1}{4}{\cos\frac{\alpha}{2}}\Big\langle(1+\vec{\sigma}_1\cdot\hat{a}_i)(\cos\frac{\alpha}{2}+\vec{\sigma}_2\cdot\hat{b}_i)
+(1-\vec{\sigma}_1\cdot\hat{a}_i)(\cos\frac{\alpha}{2}-\vec{\sigma}_2\cdot\hat{b}_i)\Big\rangle\nonumber\\
=&\dfrac{1}{2}{\cos\frac{\alpha}{2}}\Big\langle\cos\dfrac{\alpha}{2}+\vec{\sigma}_1\cdot\hat{a}_i\vec{\sigma}_2\cdot\hat{b}_i\Big\rangle,
\end{align}
 and,
  \begin{align}
{\cal P}({\bar a}_i = b_1^{(i)}=b_2^{(i)})=&\dfrac{1}{4}{\cos\frac{\alpha}{2}}\Big\langle(1-\vec{\sigma}_1\cdot\hat{a}_i)(\cos\frac{\alpha}{2}+\vec{\sigma}_2\cdot\hat{b}_i)
+(1+\vec{\sigma}_1\cdot\hat{a}_i)(\cos\frac{\alpha}{2}-\vec{\sigma}_2\cdot\hat{b}_i)\Big\rangle\nonumber\\
=&\dfrac{1}{2}{\cos\frac{\alpha}{2}}\Big\langle\cos\dfrac{\alpha}{2}-\vec{\sigma}_1\cdot\hat{a}_i\vec{\sigma}_2\cdot\hat{b}_i\Big\rangle.
\end{align}
 Imposing the  non-classicality condition,  ${\cal S}_2 < 0$,  we arrive at the inequality,
 \begin{equation}
 3\cos^2\dfrac{\alpha}{2} -\sum_{i=1}^3 \langle\vec{\sigma}_1\cdot\hat{a}_i\vec{\sigma}_2\cdot\hat{b}_i\rangle^2 <0.
 \end{equation}
  The range of $\alpha$ can be fixed by demanding all the separable states to violate this inequality. We can consider the separable state, without any loss of generality, to be $\rho_s=\dfrac{1}{4}(1+\sigma_{1z})(1+\sigma_{2z})$. For this state, the LHS of the inequality (\ref{Ineqnl}) assumes the following form,
  \begin{equation}
 3\cos^2\dfrac{\alpha}{2} - \langle a_{1z}b_{1z}\rangle^2-\langle a_{2z}b_{2z}\rangle^2-\langle a_{3z}b_{3z}\rangle^2\geq 3\cos^2\dfrac{\alpha}{2} - 1.
 \end{equation}
 The second step follows as $\hat{a}_1\perp \hat{a}_2 \perp \hat{a}_3$ and $\hat{b}_1\perp\hat{b}_2 \perp \hat{b}_3$. In order that all the separable states violate this inequality $ 1\leq  3\cos^2\dfrac{\alpha}{2} < 3$, which, in turn, implies $0< \alpha \leq \frac{\pi}{2}$.
 \subsection{Proof of non-linear entanglement inequality: III}
 \label{Entnlinapp_3}
 The sum of pseudoprobabilities for entanglement inequality is as follows:
  \begin{eqnarray}
   {\cal S}_3  & = &  \sum_{i=1}^3  \Big[\mathcal{P}(a_i = b^{(i)}_1 = b^{(i)}_2)
     +\dfrac{1}{2}\Big\{ P(a_i)\mathcal{P}(\overline{a}^{(i)}_1, \overline{a}^{(i)}_2) \nonumber \\
&+ & P(\overline{a}_i)\mathcal{P}(a^{(i)}_1, a^{(i)}_2) 
+ P(b_i)\mathcal{P}(\overline{b}^{(i)}_1, \overline{b}^{(i)}_2) \nonumber \\
&+&  P(\overline{b}_i)\mathcal{P}(b^{(i)}_1, b^{(i)}_2) 
      +  P(a_i)\mathcal{P}(\overline{b}^{(i)}_1, \overline{b}^{(i)}_2)  \nonumber \\
&+  & P(\overline{a}_i)\mathcal{P}(b^{(i)}_1, b^{(i)}_2) 
     +  \mathcal{P}(\overline{a}^{(i)}_1, \overline{a}^{(i)}_2) P(b_i) \nonumber \\
&+  & \mathcal{P}(a^{(i)}_1, a^{(i)}_2) P(\overline{b}_i)\Big\}\Big].
   \end{eqnarray}
  We next substitute the values of all the pseudoprobabilities as expectation values of corresponding PP operators as follows:
    \begin{eqnarray}
   {\cal S}_3  & = &  \sum_{i=1}^3  \dfrac{1}{2}\cos\frac{\alpha}{2}\Big(\cos\frac{\alpha}{2}+\vec{\sigma}_1\cdot\hat{a}_i\vec{\sigma}_2\cdot\hat{b}_i\Big)
     +\dfrac{1}{2}\Big\{\dfrac{1}{4}\cos\frac{\alpha}{2}(1+\vec{\sigma}_1\cdot\hat{a}_i)(\cos\dfrac{\alpha}{2}-\vec{\sigma}_1\cdot\hat{a}_i) \nonumber \\
&+ & \dfrac{1}{4}\cos\frac{\alpha}{2}(1-\vec{\sigma}_1\cdot\hat{a}_i)(\cos\dfrac{\alpha}{2}+\vec{\sigma}_1\cdot\hat{a}_i) 
+ \dfrac{1}{4}\cos\frac{\alpha}{2}(1+\vec{\sigma}_2\cdot\hat{b}_i)(\cos\dfrac{\alpha}{2}-\vec{\sigma}_2\cdot\hat{b}_i)  \nonumber \\
&+&  \dfrac{1}{4}\cos\frac{\alpha}{2}(1-\vec{\sigma}_2\cdot\hat{b}_i)(\cos\dfrac{\alpha}{2}+\vec{\sigma}_2\cdot\hat{b}_i)  
      +  \dfrac{1}{4}\cos\frac{\alpha}{2}(1+\vec{\sigma}_1\cdot\hat{a}_i)(\cos\dfrac{\alpha}{2}-\vec{\sigma}_2\cdot\hat{b}_i)    \nonumber \\
&+  & \dfrac{1}{4}\cos\frac{\alpha}{2}(1-\vec{\sigma}_2\cdot\hat{a}_i)(\cos\dfrac{\alpha}{2}+\vec{\sigma}_2\cdot\hat{b}_i)  
     + \dfrac{1}{4}\cos\frac{\alpha}{2}(\cos\dfrac{\alpha}{2}-\vec{\sigma}_1\cdot\hat{a}_i)(1+\vec{\sigma}_2\cdot\hat{b}_i)   \nonumber \\
&+  &  \dfrac{1}{4}\cos\frac{\alpha}{2}(\cos\dfrac{\alpha}{2}+\vec{\sigma}_1\cdot\hat{a}_i)(1-\vec{\sigma}_2\cdot\hat{b}_i) \Big\}\nonumber\\
&=&  \dfrac{1}{2}\cos\frac{\alpha}{2}\Big( 9\cos\dfrac{\alpha}{2}+\sum_{i=1}^3\vec{\sigma}_1\cdot\hat{a}_i\vec{\sigma}_2\cdot\hat{b}_i-\dfrac{1}{2}\langle \vec{\sigma}_1\cdot\hat{a}_i+\vec{\sigma}_2\cdot\hat{b}_i\rangle^2\Big).
   \end{eqnarray}
   Imposing the nonclassicality condition-- $   {\cal S}_3 <0$, the following entanglement inequality emerges,
   \begin{align}
   9\cos\dfrac{\alpha}{2}+\sum_{i=1}^3\Big(\vec{\sigma}_1\cdot\hat{a}_i\vec{\sigma}_2\cdot\hat{b}_i-\dfrac{1}{2}\langle \vec{\sigma}_1\cdot\hat{a}_i+\vec{\sigma}_2\cdot\hat{b}_i\rangle^2\Big)<0.
   \end{align}
   We demand that all the separable states should violate this inequality. It fixes the range of $\alpha$ to be $ 0 < \alpha \leq {\rm arccos}\Big(-\frac{79}{81}\Big)$.

 \section{Proof of condition for quantum discord}
 \label{Discord_app}
 The pseudoprobabilities are as follows:
 \begin{eqnarray}
 {\cal P}_D ^{i} &= & {\cal P}(a_i= b^{(i)}_1=b^{(i)}_2)  + 
  P(a_i) {\cal P}(\overline{b}^{(i)}_1\overline{b}^{(i)}_2) 
+ P(\overline{a}_i) {\cal P}(b^{(i)}_1 b^{(i)}_2) ; i=1, 2.\nonumber\\
 \end{eqnarray}
Note that the eigenbasis of $a_i \equiv \vec{\sigma}_1\cdot\hat{a}_i$ is the same as that of the reduced density matrix of the first subsystem. We rewrite the pseudo-probabilities in terms of expectation values of Pauli observables as below,
 \begin{align}
{\cal P}(a_i = b_1^{(i)}=b_2^{(i)})=&\dfrac{1}{4}{\cos\frac{\alpha}{2}}\Big\langle(1+\vec{\sigma}_1\cdot\hat{a}_i)(\cos\frac{\alpha}{2}+\vec{\sigma}_2\cdot\hat{b}_i)
+(1-\vec{\sigma}_1\cdot\hat{a}_i)(\cos\frac{\alpha}{2}-\vec{\sigma}_2\cdot\hat{b}_i)\Big\rangle\nonumber\\
=&\dfrac{1}{2}{\cos\frac{\alpha}{2}}\Big\langle\cos\dfrac{\alpha}{2}+\vec{\sigma}_1\cdot\hat{a}_i\vec{\sigma}_2\cdot\hat{b}_i\Big\rangle,
\end{align}
and,
\begin{align}
&P(a_i) {\cal P}(\overline{b}_1^{(i)}\overline{b}_2^{(i)})+P(\overline{a}_i) {\cal P}(b^{(i)}_1 b^{(i)}_2)\nonumber\\
=&\dfrac{1}{4}{\cos\frac{\alpha}{2}}\Big\{(1+\langle\vec{\sigma}_1\cdot\hat{a}_i\rangle)(\cos\frac{\alpha}{2}-\langle\vec{\sigma}_2\cdot\hat{b}_i\rangle)
+{\cos\frac{\alpha}{2}}(1-\langle\vec{\sigma}_1\cdot\hat{a}_i\rangle)(\cos\frac{\alpha}{2}+\langle\vec{\sigma}_2\cdot\hat{b}_i\rangle)\Big\}\nonumber\\
=&\dfrac{1}{2}{\cos\frac{\alpha}{2}}\Big(\cos\dfrac{\alpha}{2}-\langle\vec{\sigma}_1\cdot\hat{a}_i\rangle\langle\vec{\sigma}_2\cdot\hat{b}_i\rangle\Big).
\end{align}
Thus, 
\begin{eqnarray}
\label{Discordcond}
 {\cal P}_D ^{i} &= & {\cal P}(a_i= b^{(i)}_1=b^{(i)}_2)  + 
  P(a_i) {\cal P}(\overline{b}^{(i)}_1\overline{b}^{(i)}_2) 
+ P(\overline{a}_i) {\cal P}(b^{(i)}_1 b^{(i)}_2) \nonumber \\
& \equiv & \lambda\Big\{(16\lambda + 2\langle  \vec{\sigma}_{1}\cdot\hat{a}_i\vec{\sigma}_{2}\cdot\hat{b}_i\rangle 
 - 2   \langle  \vec{\sigma}_{1}\cdot\hat{a}_i\rangle\langle \vec{\sigma}_{2}\cdot\hat{b}_i\rangle\Big\},
 \end{eqnarray}
where $\lambda = \dfrac{1}{4}\cos\dfrac{\alpha}{2}.$ The two--qubit states having zero discord from ${\cal D}^{1\rightarrow 2}$ are of the following form:
\begin{align}
\label{Zerodicord}
\rho = \sum_{i}p_i|\phi_i\rangle\langle \phi_i|\otimes \rho_{2i}.
\end{align} 
Without any loss of generality, we may assume $|\phi_1\rangle = |0\rangle$ and $|\phi_2\rangle = |1\rangle$, where $|0\rangle (|1\rangle)$ are eigenvalues of $\sigma_{1z}$ with eigenvalues $+1(-1)$. Let $\rho_{2i} = \dfrac{1}{2}(1+\vec{\sigma}_2\cdot\hat{c}_i)$. Thus, it is quite evident that all the correlation terms in equation (\ref{Zerodicord}) have the form $\sigma_{1z}\vec{\sigma}_2\cdot\hat{c}_i$ and the local terms have the form $\sigma_{1z}$ and $\vec{\sigma}_2\cdot\hat{c}_i$. Thus, in equation (\ref{Discordcond}), if we choose $\hat{a}_1 =\hat{z}$ and $\hat{a}_2 = \hat{x}$ (say), only ${\cal P}^1_D$ turns negative and ${\cal P}^2_D$ is always non-negative for any value of $\alpha$ and any choice of $\hat{b}_1$ and $\hat{b}_2$ for a state of the form (\ref{Zerodicord}).  

In order to show that there exist discordant states for which both the pseudoprobabilities, ${\cal P}_D^1$ and ${\cal P}_D^2$, are negative, consider the two-qubit Werner states $\rho_W = \frac{1}{4}(1-\alpha\vec{\sigma}_1\cdot\vec{\sigma}_2)$. The non-negativity of eigenvalues of $\rho_W$ demands that $\alpha \in \Big[-\dfrac{1}{3}, 1\Big]$. For any non-zero value of $\alpha$, $\rho_W$ has nonzero discord \cite{Ollivier01}. If we choose, $\hat{a}_1 = \hat{x}, \hat{a}_2 = \hat{y}$ and 
 $\hat{b}_1 = \hat{x}, \hat{b}_2 = \hat{y}$, both the pseudoprobabilities, ${\cal P}^1_D$ and ${\cal P}_D^2$, acquire negative values for some value of $\alpha$, which proves our claim.

\end{appendices}
\section*{Author Contribution Statement}
Both authors contributed equally to this work in all respects.

%--------------------------------------------------------
%-----------------------------------------------------

\bibliographystyle{unsrt}       % APS-like style for physics
%\bibliography{bibliography}   % name your BibTeX data base

\begin{thebibliography}{10}

\bibitem{Adhikary18}
Soumik Adhikary, Sooryansh Asthana, and V.~Ravishankar.
\newblock A unified framework for non-classicality: emergence of non-locality
  and entanglement.
\newblock {\em Eur. Phys. J. D}, 74(68):68, 2020.

\bibitem{Asthana20}
Sooryansh Asthana, Soumik Adhikary, and V~Ravishankar.
\newblock Non-locality and entanglement in multi-qubit systems from a unified
  framework.
\newblock {\em Quantum Information Processing}, 20(1):1--33, 2021.

\bibitem{Schrodinger26}
Ervin Schr\"{o}dinger.
\newblock Quantisation as a problem of proper values (part iv).
\newblock {\em Annalen der Physik}, 81:102--123, 1926.

\bibitem{Heisenberg25}
Werner Heisenberg.
\newblock {\em Zeitschrift f\"{u}r physik}, 33:879--893, 1925.

\bibitem{Dirac27}
P~A~M Dirac.
\newblock The physical interpretation of quantum dynamics.
\newblock {\em Proceedings of Royal Society of London A}, 113:621--641, 1927.

\bibitem{Feynman48}
R.~P. Feynman.
\newblock Space time approach to non relativistic quantum mechanics.
\newblock {\em Rev. Mod. Phys}, 20:367--387, 1948.

\bibitem{Bohm52}
D.~Bohm.
\newblock A suggested interpretation of quantum theory in terms of hidden
  variables.
\newblock {\em Phys. Rev.}, 85:166--179, 1952.

\bibitem{Nelson66}
E.~Nelson.
\newblock Derivation of the schrodinger equation from newtonian mechanics.
\newblock {\em Phys. Rev.}, 150:1079, Oct 1966.

\bibitem{Einstein35}
A.~Einstein, B.~Podolsky, and N.~Rosen.
\newblock Can quantum-mechanical description of physical reality be considered
  complete?
\newblock {\em Phys. Rev.}, 47:777, May 1935.

\bibitem{Scrodinger35}
E.~Schr\"{o}dinger.
\newblock {Die gegenw\"{a}rtige Situation in der Quantenmechanik}.
\newblock {\em Naturwissenschaften}, 23(50):844, December 1935.

\bibitem{Bell64}
J.~S. Bell.
\newblock On the einstein podolsky rosen paradox.
\newblock {\em Physics Physique Fizika}, 1:195--200, Nov 1964.

\bibitem{Kochen67}
Simon Kochen and Ernst~P Specker.
\newblock The problem of hidden variables in quantum mechanics.
\newblock {\em Journal of mathematics and mechanics}, 17(1):59--87, 1967.

\bibitem{Aharonov88}
Yakir Aharonov, David~Z. Albert, and Lev Vaidman.
\newblock How the result of a measurement of a component of the spin of a
  spin-1/2 particle can turn out to be 100.
\newblock {\em Phys. Rev. Lett.}, 60:1351--1354, Apr 1988.

\bibitem{Vaidman09}
L.~Vaidman.
\newblock {\em Compendium of quantum physics: concepts, experiments, history
  and philosophy (edited by D. Greenberger, K. Hentschel and F. Weintert)}.
\newblock Springer Science \& Business Media, 2009.

\bibitem{Horodecki09}
Ryszard Horodecki, Pawe\l{} Horodecki, Micha\l{} Horodecki, and Karol
  Horodecki.
\newblock Quantum entanglement.
\newblock {\em Rev. Mod. Phys.}, 81:865--942, Jun 2009.

\bibitem{Ollivier01}
Harold Ollivier and Wojciech~H. Zurek.
\newblock Quantum discord: A measure of the quantumness of correlations.
\newblock {\em Phys. Rev. Lett.}, 88:017901, Dec 2001.

\bibitem{Knill98}
E.~Knill and R.~Laflamme.
\newblock Power of one bit of quantum information.
\newblock {\em Phys. Rev. Lett.}, 81:5672--5675, Dec 1998.

\bibitem{Shi17}
Hai-Long Shi, Si-Yuan Liu, Xiao-Hui Wang, Wen-Li Yang, Zhan-Ying Yang, and Heng
  Fan.
\newblock Coherence depletion in the grover quantum search algorithm.
\newblock {\em Phys. Rev. A}, 95:032307, Mar 2017.

\bibitem{Bennett92}
Charles~H. Bennett and Stephen~J. Wiesner.
\newblock Communication via one- and two-particle operators on
  einstein-podolsky-rosen states.
\newblock {\em Phys. Rev. Lett.}, 69:2881--2884, Nov 1992.

\bibitem{Bennett93}
Charles~H. Bennett, Gilles Brassard, Claude Cr\'epeau, Richard Jozsa, Asher
  Peres, and William~K. Wootters.
\newblock Teleporting an unknown quantum state via dual classical and
  einstein-podolsky-rosen channels.
\newblock {\em Phys. Rev. Lett.}, 70:1895--1899, Mar 1993.

\bibitem{Ekert91}
Artur~K. Ekert.
\newblock Quantum cryptography based on bell's theorem.
\newblock {\em Phys. Rev. Lett.}, 67:661--663, Aug 1991.

\bibitem{Streltsov12}
Alexander Streltsov, Hermann Kampermann, and Dagmar Bru\ss{}.
\newblock Quantum cost for sending entanglement.
\newblock {\em Phys. Rev. Lett.}, 108:250501, Jun 2012.

\bibitem{Kanjilal18}
Som Kanjilal, Aiman Khan, C.~Jebarathinam, and Dipankar Home.
\newblock Remote state preparation using correlations beyond discord.
\newblock {\em Phys. Rev. A}, 98:062320, Dec 2018.

\bibitem{Birkhoff36}
Garrett Birkhoff and John~Von Neumann.
\newblock The logic of quantum mechanics.
\newblock {\em Annals of Mathematics}, 37(4):823--843, 1936.

\bibitem{Fine82}
Arthur Fine.
\newblock Hidden variables, joint probability, and the bell inequalities.
\newblock {\em Phys. Rev. Lett.}, 48:291, Feb 1982.

\bibitem{Hill61}
H.~Margenau and Robert~N. Hill.
\newblock {Correlation between Measurements in Quantum Theory: }.
\newblock {\em Progress of Theoretical Physics}, 26(5):722--738, 11 1961.

\bibitem{Barut88}
A.~O. Barut, M.~Bo{\v{z}}i{\'{c}}, and Z.~Mari{\'{c}}.
\newblock Joint probabilities of noncommuting operators and incompleteness of
  quantum mechanics.
\newblock {\em Foundations of Physics}, 18(10):999--1012, Oct 1988.

\bibitem{Peres96}
Asher Peres.
\newblock Separability criterion for density matrices.
\newblock {\em Phys. Rev. Lett.}, 77:1413--1415, Aug 1996.

\bibitem{Horodecki96a}
Michał Horodecki, Paweł Horodecki, and Ryszard Horodecki.
\newblock Separability of mixed states: necessary and sufficient conditions.
\newblock {\em Physics Letters A}, 223(1):1 -- 8, 1996.

\bibitem{Guhne03}
O.~G\"uhne, P.~Hyllus, D.~Bruss, A.~Ekert, M.~Lewenstein, C.~Macchiavello, and
  A.~Sanpera.
\newblock Experimental detection of entanglement via witness operators and
  local measurements.
\newblock {\em Journal of Modern Optics}, 50(6-7):1079--1102, 2003.

\bibitem{Vaidman08}
L.~Vaidman.
\newblock Protective measurements.
\newblock {\em arXiv}, 0801.2761, 2008.

\bibitem{Pusey14}
Matthew~F. Pusey.
\newblock Anomalous weak values are proofs of contextuality.
\newblock {\em Phys. Rev. Lett.}, 113:200401, Nov 2014.

\bibitem{Hosoya10}
Akio Hosoya and Yutaka Shikano.
\newblock Strange weak values.
\newblock {\em Journal of Physics A: Mathematical and Theoretical},
  43(38):385307, aug 2010.

\bibitem{Lundeen12}
Jeff~S. Lundeen and Charles Bamber.
\newblock Procedure for direct measurement of general quantum states using weak
  measurement.
\newblock {\em Phys. Rev. Lett.}, 108:070402, Feb 2012.

\bibitem{Bamber14}
Charles Bamber and Jeff~S. Lundeen.
\newblock Observing dirac's classical phase space analog to the quantum state.
\newblock {\em Phys. Rev. Lett.}, 112:070405, Feb 2014.

\bibitem{Higgins15}
B.~L. Higgins, M.~S. Palsson, G.~Y. Xiang, H.~M. Wiseman, and G.~J. Pryde.
\newblock Using weak values to experimentally determine ``negative
  probabilities'' in a two-photon state with bell correlations.
\newblock {\em Phys. Rev. A}, 91:012113, Jan 2015.

\bibitem{Piacentini16}
F.~Piacentini, A.~Avella, M.~P. Levi, M.~Gramegna, G.~Brida, I.~P. Degiovanni,
  E.~Cohen, R.~Lussana, F.~Villa, A.~Tosi, F.~Zappa, and M.~Genovese.
\newblock Measuring incompatible observables by exploiting sequential weak
  values.
\newblock {\em Phys. Rev. Lett.}, 117:170402, Oct 2016.

\bibitem{Dziewior19}
Jan Dziewior, Lukas Knips, Demitry Farfurnik, Katharina Senkalla, Nimrod
  Benshalom, Jonathan Efroni, Jasmin Meinecke, Shimshon Bar-Ad, Harald
  Weinfurter, and Lev Vaidman.
\newblock Universality of local weak interactions and its application for
  interferometric alignment.
\newblock {\em Proceedings of the National Academy of Sciences},
  116(8):2881--2890, 2019.

\bibitem{Dirac42}
P.~A.~M. Dirac.
\newblock Bakerian lecture. the physical interpretation of quantum mechanics.
\newblock {\em Proceedings of the Royal Society of London A: Mathematical,
  Physical and Engineering Sciences}, 180(980):1--40, 1942.

\bibitem{Feynman87}
Richard~P Feynman.
\newblock {\em Chapter- 13, Quantum implications: essays in honour of David
  Bohm (edited by B. Hiley and F. Peat)}.
\newblock Taylor \& Francis, 2012.

\bibitem{Chandler92}
C~Chandler, L~Cohen, C~Lee, M~Scully, and K~Wodkiewicz.
\newblock Quasi-probability distribution for spin-1/2 particles.
\newblock {\em Foundations of physics}, 22(7):867--878, 1992.

\bibitem{Adhikary17}
Soumik Adhikary and V.~Ravishankar.
\newblock Quantum logic and non-classicality: an explicit formulation.
\newblock {\em arXiv}, 1710.04371v1, 2017.

\bibitem{Weyl27}
H.~Weyl.
\newblock Quantenmechanik und gruppentheorie.
\newblock {\em Zeitschrift f{\"u}r Physik}, 46(1):1--46, Nov 1927.

\bibitem{Popescu94}
Sandu Popescu and Daniel Rohrlich.
\newblock Quantum nonlocality as an axiom.
\newblock {\em Foundations of Physics}, 24(3):379--385, 1994.

\bibitem{Wiseman02}
H.~M. Wiseman.
\newblock Weak values, quantum trajectories, and the cavity-qed experiment on
  wave-particle correlation.
\newblock {\em Phys. Rev. A}, 65:032111, Feb 2002.

\bibitem{Vaidman17}
Lev Vaidman, Alon Ben-Israel, Jan Dziewior, Lukas Knips, Mira Wei\ss{}l, Jasmin
  Meinecke, Christian Schwemmer, Ran Ber, and Harald Weinfurter.
\newblock Weak value beyond conditional expectation value of the pointer
  readings.
\newblock {\em Phys. Rev. A}, 96:032114, Sep 2017.

\bibitem{Kirkwood33}
John~G. Kirkwood.
\newblock Quantum statistics of almost classical assemblies.
\newblock {\em Phys. Rev.}, 44:31--37, Jul 1933.

\bibitem{Barut57}
A.~O. Barut.
\newblock Distribution functions for noncommuting operators.
\newblock {\em Phys. Rev.}, 108:565--569, Nov 1957.

\bibitem{Johansen_04}
Lars~M. Johansen and Alfredo Luis.
\newblock Nonclassicality in weak measurements.
\newblock {\em Phys. Rev. A}, 70:052115, Nov 2004.

\bibitem{Johansen__04}
Lars~M. Johansen.
\newblock Nonclassical properties of coherent states.
\newblock {\em Physics Letters A}, 329(3):184–187, Aug 2004.

\bibitem{Pan20}
A.~K. Pan.
\newblock Interference experiment, anomalous weak value, and leggett-garg test
  of macrorealism.
\newblock {\em Phys. Rev. A}, 102:032206, Sep 2020.

\bibitem{Streltsov17}
Alexander Streltsov, Gerardo Adesso, and Martin~B. Plenio.
\newblock Colloquium: Quantum coherence as a resource.
\newblock {\em Rev. Mod. Phys.}, 89:041003, Oct 2017.

\bibitem{Aharonov16}
Yakir Aharonov, Fabrizio Colombo, Sandu Popescu, Irene Sabadini, Daniele~C.
  Struppa, and Jeff Tollaksen.
\newblock Quantum violation of the pigeonhole principle and the nature of
  quantum correlations.
\newblock {\em Proceedings of the National Academy of Sciences},
  113(3):532--535, 2016.

\bibitem{Reznik20}
Gregory Reznik, Shrobona Bagchi, Justin Dressel, and Lev Vaidman.
\newblock Footprints of quantum pigeons.
\newblock {\em Phys. Rev. Research}, 2:023004, Apr 2020.

\bibitem{Hardy93}
Lucien Hardy.
\newblock Nonlocality for two particles without inequalities for almost all
  entangled states.
\newblock {\em Phys. Rev. Lett.}, 71:1665--1668, Sep 1993.

\bibitem{Aharonov08}
Yakir Aharonov and Daniel Rohrlich.
\newblock {\em Quantum paradoxes: quantum theory for the perplexed}.
\newblock John Wiley \& Sons, 2008.

\bibitem{Guhne04}
Otfried G\"uhne.
\newblock Characterizing entanglement via uncertainty relations.
\newblock {\em Phys. Rev. Lett.}, 92:117903, Mar 2004.

\end{thebibliography}

\end{document}